\documentclass[%
 showpacs,
 showkeys,
 amsmath,amssymb,
prx,
]{revtex4-1}

\usepackage{graphicx}
\usepackage{dcolumn}
\usepackage{bm}

\usepackage{CJK}

\usepackage{color}
\usepackage{amssymb}
\usepackage{amsfonts}

\usepackage[colorlinks,urlcolor=blue,linkcolor=blue,citecolor=blue,anchorcolor=blue]{hyperref}

\begin{document}


\preprint{APS/123-QED}

\title{Type-II Dirac photonic lattices}

\author{Kaichao Jin$^{1,2,\dag}$}
\author{Hua Zhong$^{1,2,\dag}$}
\author{Yongdong Li$^1$}
\author{Fangwei Ye$^{3}$}
\author{Yanpeng Zhang$^{1}$}
\author{Fuli Li$^4$}
\author{Chunliang Liu$^1$}
\author{Yiqi Zhang$^{1,2,}$}
\email{zhangyiqi@mail.xjtu.edu.cn}
\affiliation{%
 $^1$Key Laboratory for Physical Electronics and Devices of the Ministry of Education,
Xi'an Jiaotong University, Xi'an 710049, China \\
$^2$Guangdong Xi'an Jiaotong University Academy, Foshan 528300, China \\
$^3$School of Physics and Astronomy, Shanghai Jiao Tong University, Shanghai 200240, China \\
$^4$Department of Applied Physics, School of Science, Xi'an Jiaotong University, Xi'an 710049, China\\
$^\dag$These authors contribute equally to this work.
}%

\date{\today}

\begin{abstract}
  \noindent Different from the Fermi surface of the type-I Dirac semimetal being a point,
  that of the type-II Dirac semimetal is a pair of crossing lines because the
  Dirac cone is tilted with open and hyperbolic isofrequency contours.
  As an optical analogy, type-II Dirac photonic lattices have been also designed.
  Here, we report type-II Dirac cones in Lieb-like photonic lattices composed of identical waveguide channels,
  and the anisotropy of the band structure is due to neither the refractive index change nor the environment,
  but only the spatial symmetry of the lattice; therefore, the proposal is advantage and benefits experimental observation.
  Conical diffractions and Klein tunneling in the parametric type-II photonic lattice are investigated in detail.
  Our results provide a simple and experimental feasible platform to study two-dimensional topological photonic
  and other nonrelativistic phenomena around type-II Dirac cones.
\end{abstract}

\keywords{type-II Dirac cone, asymmetric conical diffraction, Klein tunneling}
\maketitle

%
\section{Introduction}
If a photonic lattice possesses Dirac cones \cite{leykam.aipx.1.101.2016} in its band structure,
it can be called a Dirac photonic lattice.
One of the most famous Dirac photonic lattices is the photonic graphene \cite{zandbergen.prl.104.043903.2010,rechtsman.prl.111.103901.2013,rechtsman.np.7.153.2013,crespi.njp.15.013012.2013,zeuner.ol.39.602.2014,plotnik.nm.13.57.2014,song.nc.6.6272.2015,nalitov.prl.114.026803.2015},
which is also known as the honeycomb lattice.
In addition, the Lieb lattice \cite{taie.sa.1.1500854.2015,vicencio.prl.114.245503.2015,mukherjee.prl.114.245504.2015,diebel.prl.116.183902.2016,ozawa.prl.118.175301.2017,xia.prl.121.263902.2018},
the kagome lattice \cite{hassan.np.13.697.2019,li.np.2019},
the superhoneycomb lattices \cite{zhong.adp.529.1600258.2017,kang.adp.531.1900295.2019},
and some others \cite{zhong.oe.27.6300.2019}
can be also classified into Dirac photonic lattices.
It is worth mentioning that the Dirac cones of these photonic lattices aforementioned are mostly type-I,
even though some of them are tilted.
Actually, there are also type-II and type-III Dirac cones \cite{milicevic.prx.9.031010.2019}.
The main difference between them is that the Fermi surface of the type-I Dirac cone is a point,
that of the type-II Dirac cone is a pair of crossing lines, and that of the type-III Dirac cone is a line.
In other words, the isofrequency contours of the type-I Dirac cones are closed,
while those of the type-II counterparts are open and have hyperbolic profiles (e.g., on both $+k_y$ and $-k_y$ directions).
The isofrequency contours of the type-III ones are also open,
but only have hyperbolic profiles on one side (e.g., on either $+k_y$ or $-k_y$ direction).
Note that there is also a different definition on the type-III degeneracy \cite{li.arxiv}.
As a plethora of efforts being implemented to explore the
type-II Weyl-like features \cite{soluyanov.nature.527.495.2015,xu.prl.115.265304.2015,chen.nc.7.13038.2016,autes.prl.117.066402.2016,noh.np.13.611.2017,chen.prb.97.155152.2018,xie.prl.122.104302.2019}
both in condensed matter physics and photonics,
type-II Dirac objects attract ongoing attentions too \cite{yan.nc.8.257.2017,noh.prl.119.016401.2017,chang.prl.119.026404.2017,fei.am.30.1801556.2018,politano.prl.121.086804.2018,kim.prm.2.104203.2018,wang.prb.98.115164.2018},
especially in photonics \cite{lin.prb.96.075438.2017,pyrialakos.prl.119.113901.2017,wang.npjqm.2.54.2017,hu.prl.121.024301.2018,mann.nc.9.2194.2018,milicevic.prx.9.031010.2019}.
Different from ubiquitous type-I Dirac cones, appearance of type-II Dirac cones demands either high anisotropic lattice arrangement or accurate manipulation of the lattice environment.
Despite the difference, quasiparticles corresponding to both the two types of Dirac cones are massless, and can be described by the massless Dirac Hamiltonian \cite{li.prb.96.081106.2017},
thereby Dirac materials become unique paradigms to explore relativistic Dirac-related phenomena, such as Klein tunneling \cite{katsnelson.np.2.620.2006}.

Being a relativistic phenomenon, Klein tunneling means that a massless particle can overpass a barrier higher than its energy freely.
Surprisingly, this nonintuitive prediction was also explained successfully in the frame of classical electromagnetic theory
and simulated and observed in classical systems, such as deformed honeycomb lattice \cite{bahat-treidel.prl.104.063901.2010,bahat-treidel.pra.84.021802.2011},
waveguide superlattices \cite{longhi.prb.81.075102.2010,dreisow.epl.97.10008.2012}, matamaterials \cite{sun.sr.7.9678.2017}, and pseudospin-1 photonic crystals \cite{fang.research.2019.3054062.2019}.
As far as we know, Klein tunneling in type-II Dirac photonic lattices is still open to be explored.

\begin{figure}
  \includegraphics*[width=0.5\columnwidth]{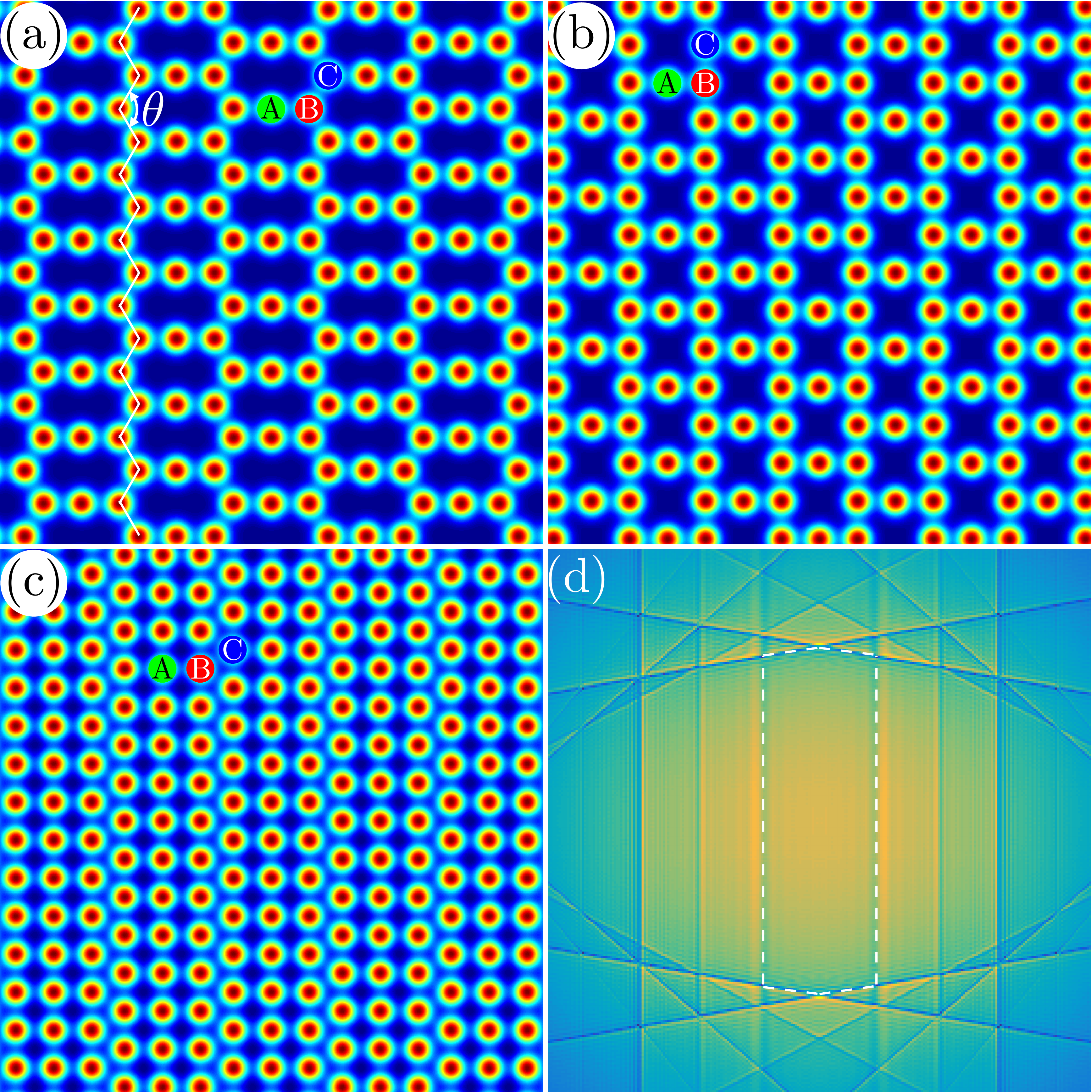}
  \caption{(a) Lattice with a controllable parameter $\theta$. Here $\theta=2\pi/3$.
  (b) Lattice with $\theta=\pi$, which is also the dislocated Lieb lattice.
  (c) Lattice with $\theta=\pi/3$.
  (d) Far-field diffraction pattern of the lattice in (c). The first Brillouin zone is shown by the dashed deformed hexagon.
  Unit cell of the lattice includes three sites that are indicated by green, red and blue colors with letters A, B and C.
  Other parameters: $a=1.4$, $p=10$ and $d=0.5$.}
  \label{fig1}
\end{figure}

In this work, we unveil type-II Dirac cones in novel but simple lattice waveguide arrays constructed by
identical waveguide channels that are transversely arranged in morphable lattice profiles with three sites in one unit cell,
via a controllable angle parameter $\theta$ in the range $[\pi/6,\pi]$ (see Figure \ref{fig1}).
When the angle reaches its supreme value, a dislocated Lieb lattice \cite{li.prb.97.081103.2018} that has both type-I and type-III Dirac cones \cite{milicevic.prx.9.031010.2019} is created.
With the angle value decreases, the lattice deforms with the type-III Dirac cones reducing into tilted type-I Dirac cones.
Further deformation of the lattice by minishing the angle value, the type-I Dirac cones tilt even further
and evolve into the type-II Dirac cones gradually.
Light modulation and manipulation based on these various Dirac cones are elucidated by the corresponding conical diffractions,
and for the first time, the Klein tunneling is demonstrated in the type-II Dirac photonic lattice.
Our results provide a new design for type-II Dirac photonic lattices,
which has advantage of simplicity since the appearance of the type-II Dirac cones is only dependent on the spatial symmetry of the lattice.

\section{Theoretical Modelling}
Propagation of a light beam in a photonic lattice can be faithfully described by the Schr\"odinger-like paraxial wave equation:
\begin{equation}\label{eq1}
  i \frac{\partial \psi}{\partial z}=
  -\frac{1}{2}\left(\frac{\partial^2}{\partial x^2}+\frac{\partial^2}{\partial {y}^{2}}\right) \psi
  -\mathcal{R}(x, y) \psi,
\end{equation}
where the transverse $(x,y)$ and longitudinal $z$ coordinates are
normalized to the characteristic transverse scale $r_0$ and the diffraction
length $L_{\rm dif}=kr_0^2$, respectively; $k = 2\pi n_0/\lambda$ is the
wavenumber with $n_0$ being the background refractive index and $\lambda$ the wavelength.
The lattice potential $\mathcal{R}(x, y)=p \sum_{n, m} \mathcal{Q}(x-x_{n}, y-y_{m})$
is composed of Gaussian waveguides $\mathcal{Q}=\exp(-x^{2} / a_{x}^{2}-y^{2} / a_{y}^{2})$
with $p$ being the depths of two sublattices, $a_{x,y}$ waveguide widths,
and $(x_n,y_m)$ the transverse location of each waveguide channel.
The lattice constant is labeled as $d$ which is the distance between two nearest-neighbor sites.
The photonic lattice can be prepared by using, for instance,
the femtosecond laser writing technique in fused silica material \cite{rechtsman.nature.496.196.2013,stuzer.nature.560.461.2019,lustig.nature.567.356.2019}
and the optically induced technique in photorefractive crystals \cite{song.nc.6.6272.2015,xia.prl.121.263902.2018,wang.nature.577.42.2020}.
Taking the former technique as an example, one can use the following parameters:
$\lambda=633\,\rm nm$,
$d=  15\,\mu \rm m$,
$a_x = a_y =  10\,\mu \rm m$.
If we choose the transverse scale $r_0=  10\,\mu\rm m$,
the diffraction length $L_{\rm dif}=1.4\,\rm mm$.
The lattice depth $p=10$ corresponds to refractive index change
of $7 \times 10^{-4}$ in a real physical system.

\begin{figure}
  \includegraphics[width=0.5\columnwidth]{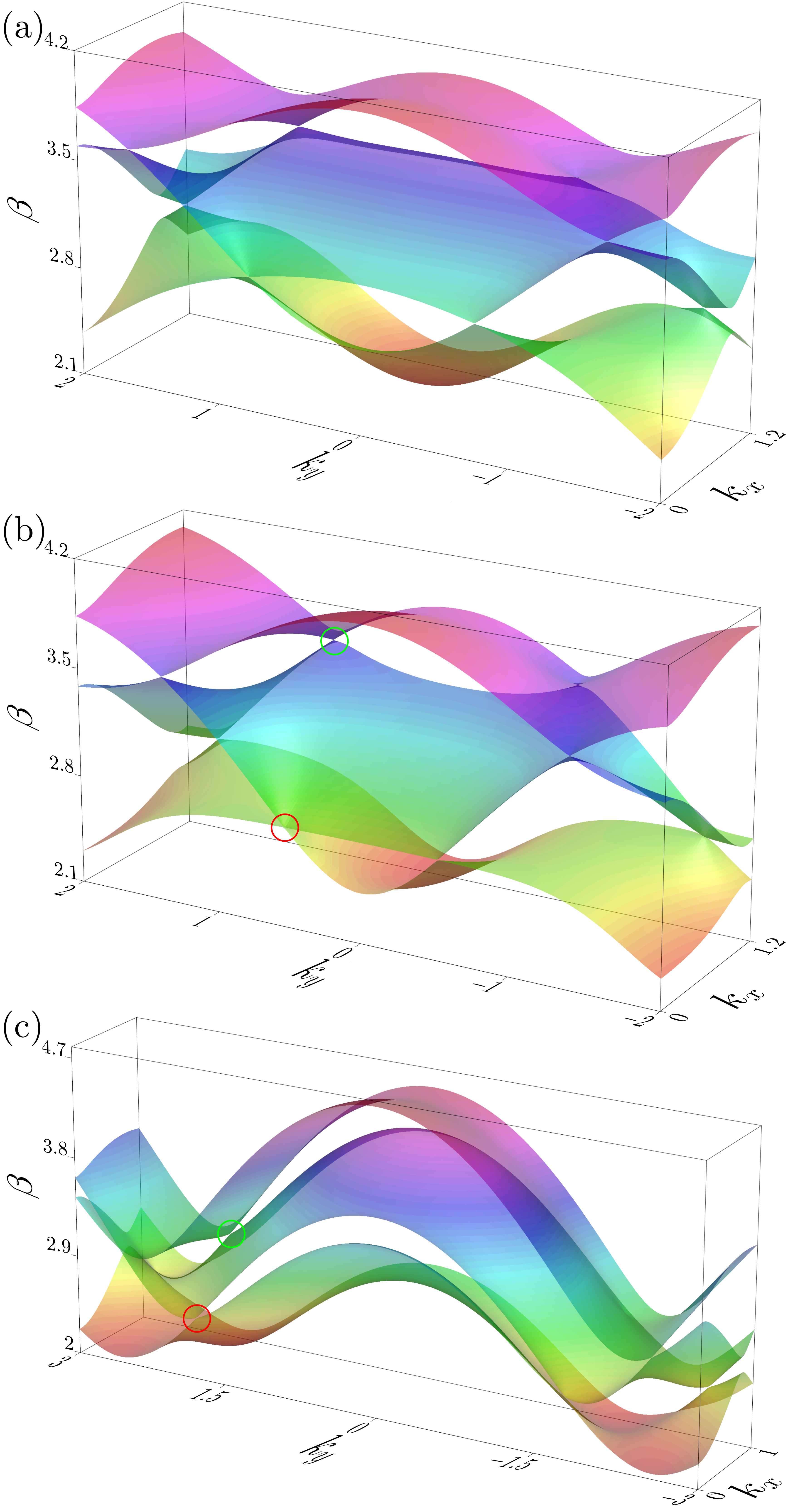}
  \caption{(a)-(c) Two-dimensional band structures corresponding to Figure \ref{fig1}(a)-\ref{fig1}(c).}
  \label{fig2}
\end{figure}

The solution of Equation (\ref{eq1}) can be written as
$\psi(x,y,z)=u(x,y)\exp(i\beta z)$ with $\beta$ being the propagation constant (or ``energy'' of the quasiparticle)
and $u(x,y)$ the Bloch mode.
Plugging this solution into Equation (\ref{eq1}), one obtains
\begin{equation}\label{eq2}
\beta u = \frac{1}{2}\left(\frac{\partial^2}{\partial x^2}+\frac{\partial^2}{\partial {y}^{2}}\right) u + \mathcal{R}(x,y) u,
\end{equation}
which can be solved numerically by using the plane-wave expansion method.
As a periodic function of Bloch momenta $k_{x,y}$,
$\beta(k_x,k_y)$ is the band structure of the photonic lattice ${\mathcal R}(x,y)$.
Clearly, photonic lattices with different geometries possess different band structures \cite{tarruell.nature.483.302.2012}.
One may see the lattice transverse profile shown in Figure \ref{fig1}(a),
which possesses three sites (labeled as A, B and C) in one unit cell.
Now, we introduce another controlling parameter,
the angle $\theta$ between sites B and C along vertical $y$ direction,
as shown by the zigzag line.
Here in Figure \ref{fig1}(a), the angle $\theta=2\pi/3$.
If the angle $\theta=\pi$, one obtains the dislocated Lieb lattice \cite{li.prb.97.081103.2018},
as shown in Figure \ref{fig1}(b), which is
different from the traditional Lieb lattice \cite{vicencio.prl.114.245503.2015,mukherjee.prl.114.245504.2015}.
One can change the angle continuously to deform the lattice, and
in Figure \ref{fig1}(c), the lattice with $\theta=\pi/3$ is shown.
It is convenient to check the far-field diffraction patterns \cite{bartal.prl.94.163902.2005} of the lattices,
which can show the corresponding Brillouin zones directly.
According to numerical simulations,
one may find that the first Brillouin zones are deformed hexagons,
and the smaller the angle, the larger the deformation of the hexagons.
In Figure \ref{fig1}(d), we only show the far-field diffraction pattern of the lattice shown in Figure \ref{fig1}(c).
Since the first Brillouin zone is always a hexagon,
there must be always three sites in one unit cell no matter what the value of the angle.
Even though there are still three sites in one unit cell in the lattice with $\theta=\pi/3$,
the property may change greatly in comparison with those with $\theta=2\pi/3$ and $\theta=\pi$.
An explicit fact is that under the tight-binding approximation,
there are 2, 3 and 3 nearest-neighbor sites for sites A, B and C, respectively,
both in Figure \ref{fig1}(a) and \ref{fig1}(b).
However in Figure \ref{fig1}(c), the numbers are 4, 5 and 5.
According to the geometry of the lattice in Figure \ref{fig1}(c),
the locations of the six corners of the first Brillouin zone can be obtained theoretically
\[
\left(0,\pm\frac{(20+8\sqrt{3})\pi}{({19+8\sqrt{3}})a} \right)
\textrm{and}
\left(\pm \frac{2\pi}{(4+\sqrt{3})a},\pm\frac{(18+8\sqrt{3})\pi}{({19+8\sqrt{3}})a}\right)
\]
which are in accordance with the numerical results in Figure \ref{fig1}(d).

\begin{figure}
  \includegraphics[width=0.5\columnwidth]{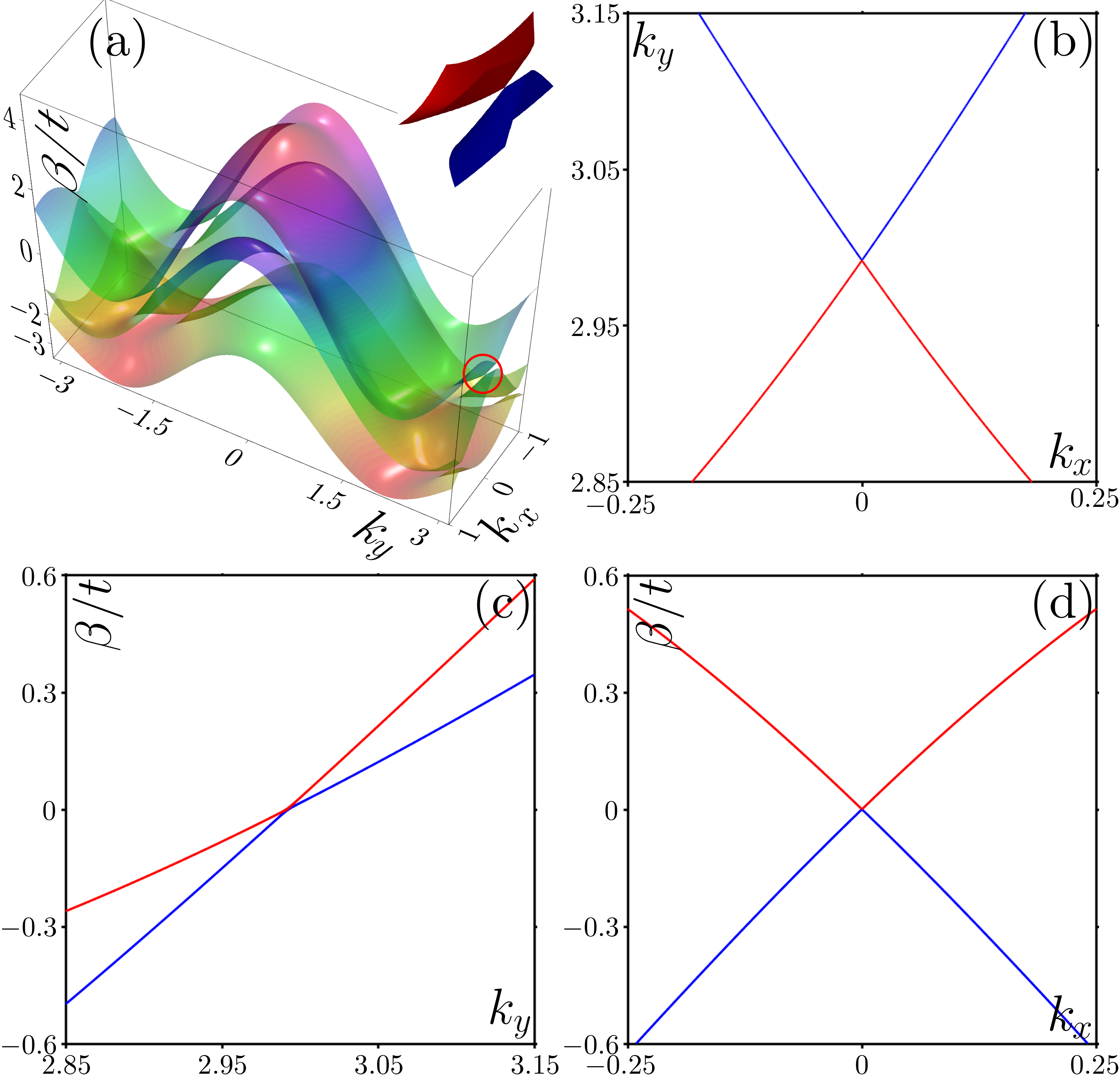}
  \caption{(a) Band structure according to Equation (\ref{eq2}). The red circle labels the type-II Dirac point that is investigated,
  and its corresponding magnification is displayed in the inset.
  In the inset, the top and middle bands are in red and blue colors, respectively.
  (b) Cross section of the Dirac cone in the $(k_x,k_y)$ plane with $\beta=0$.
  (c) Cross section of the Dirac cone in the $(k_y,\beta)$ plane with $k_x=0$.
  (d) Cross section of the Dirac cone in the $(k_x,\beta)$ plane with $k_y=4\pi/3a$.
  In (b)-(d), colors of the lines are in accordance with those of the bands in the inset in (a).}
  \label{fig3}
\end{figure}

By solving Equation (\ref{eq2}) numerically, we obtain the band structures for
the lattices in Figure \ref{fig1}, which are displayed in Figure \ref{fig2}.
In Figure \ref{fig2}(a) that corresponds to the lattice with $\theta=2\pi/3$ in Figure \ref{fig1}(a),
all the Dirac cones are tilted type-I.
While in Figure \ref{fig2}(b) that corresponds to Figure \ref{fig1}(b),
the Dirac cones between the top and bottom bands are tilted type-I,
but those between the middle and bottom bands are type-III.
When the angle decreases successively to $\theta=\pi/3$, as shown in Figure \ref{fig2}(c),
all the Dirac cones become type-II, because
the group velocity along $y$ (i.e., $-d\beta/dk_y$) of the mode around the Dirac points
does not change sign.

To understand the emergence of the type-II Dirac cones,
we theoretically analyze the couplings among sites in Figure \ref{fig1}(c)
by adopting the tight-binding approximation method and only
considering the nearest-neighbor interaction.
The corresponding Hamiltonian can be written as
\begin{equation}\label{eq3}
  {\mathcal H}=t
  \begin{bmatrix}
    2\cos({\bf k}\cdot {\bf e}_2) & e^{i{\bf k}\cdot {\bf e}_1} & e^{-i{\bf k}\cdot {\bf e}_1} \\
    e^{-i{\bf k}\cdot {\bf e}_1} & 2\cos({\bf k}\cdot {\bf e}_2) & e^{i{\bf k}\cdot {\bf e}_3}+e^{i{\bf k}\cdot {\bf e}_4} \\
    e^{i{\bf k}\cdot {\bf e}_1} & e^{-i{\bf k}\cdot {\bf e}_3}+e^{-i{\bf k}\cdot {\bf e}_4} & 2\cos({\bf k}\cdot {\bf e}_2)
  \end{bmatrix},
\end{equation}
in which ${\bf k}=[k_x,k_y]$, ${\bf e}_1=[a,0]$, ${\bf e}_2=[0,a]$,
${\bf e}_3=[\sqrt{3}a/2,a/2]$, ${\bf e}_4=[\sqrt{3}a/2,-a/2]$, and $t$ the coupling strength.
The eigenvalues of Equation (\ref{eq3}) are the band structure,
but the analytical solution is hard to obtain.
Even so, one still can obtain the locations of the Dirac points,
which are $[0,\pm4\pi/3a]$ and $[\pm2\pi/(4+\sqrt{3})a,\pm2\pi/3a]$ between the top and middle bands,
and $[0,\pm2\pi/3a]$ and $[\pm2\pi/(4+\sqrt{3})a,\pm4\pi/3a]$ between the middle and bottom bands.
Numerical band structure is displayed in Figure \ref{fig3}(a),
which looks visually the same with that in Figure \ref{fig2}(c) although quantitatively there are difference in the value of $\beta$.
Since all the Dirac cones are type-II for this case, we take the Dirac point at $[0,4\pi/3a]$ as an example without loss of generality,
and this Dirac cone is shown in the inset of Figure \ref{fig3}(a).
Before going into a subtle theoretical analysis on this type-II Dirac cone,
it is necessary to have a look at the corresponding cross sections
in the $(k_x,k_y)$ plane with $\beta=0$ [Figure \ref{fig3}(b)],
the $(k_y,\beta)$ plane with $k_x=0$ [Figure \ref{fig3}(c)],
and the $(k_x,\beta)$ plane with $k_y=4\pi/3a$ [Figure \ref{fig3}(d)].
In the $\beta=0$ plane, the intersection of the top and middle bands, as shown in Figure \ref{fig3}(b),
exhibits two crossing lines \cite{soluyanov.nature.527.495.2015,milicevic.prx.9.031010.2019}.
In Figure \ref{fig3}(c), the red and blue lines show the profile of the Dirac cone in the cross section $k_x=0$
which clearly elucidates that the sign of the slope $d\beta/dk_y$ does not change along $k_y$ direction.
While in Figure \ref{fig3}(d), the profile of the Dirac cone
in the cross section $k_y=0$ is symmetric about $k_x=0$ (i.e., sign of $d\beta/dk_x$ changes),
which is similar to that for a (tilted) type-I Dirac points.
To this regard,
if we expand the Hamiltonian in the infinitesimal region $(p_x=k_x-k_x^{\rm D},p_y=k_y-k_y^{\rm D})$
around the Dirac point $(k_x^{\rm D},k_y^{\rm D})$,
we can only consider the component along $k_y$ direction and let $p_x=0$ safely.
As a result, the corresponding Hamiltonian can be written as
\begin{equation}\label{eq4}
{\mathcal H}=t
\begin{bmatrix}
 \sqrt{3} a p_y-1 & 1 & 1 \\
 1 & \sqrt{3} a p_y-1 & -\frac{\sqrt{3}}{2}  a p_y-1 \\
 1 & -\frac{\sqrt{3}}{2}  a p_y-1 & \sqrt{3} a p_y-1
\end{bmatrix}.
\end{equation}
The eigenvalues of this Hamiltonian are
\begin{equation}\label{eq5}
  \beta_1 = \frac{3}{2} \sqrt{3} a t p_y,\,
  \beta_2 = \frac{5}{6} \sqrt{3} a t p_y,\,
  \beta_3 = \frac{2}{3} \sqrt{3} a t p_y -3t.
\end{equation}
Clearly, one may find the relation $\beta_1|_{p_y=0}=\beta_2|_{p_y=0} = 0$,
which indicates that the two bands are degenerated at the point $(p_x=0,p_y=0)$ ---
the location of the Dirac point.
One also finds that $d\beta_1/dp_y$ and $d\beta_2/dp_y$ always have the same sign,
therefore the velocity does not change its sign around the Dirac point,
and the Dirac cone is type-II definitely.

\begin{figure}
  \includegraphics[width=0.5\columnwidth]{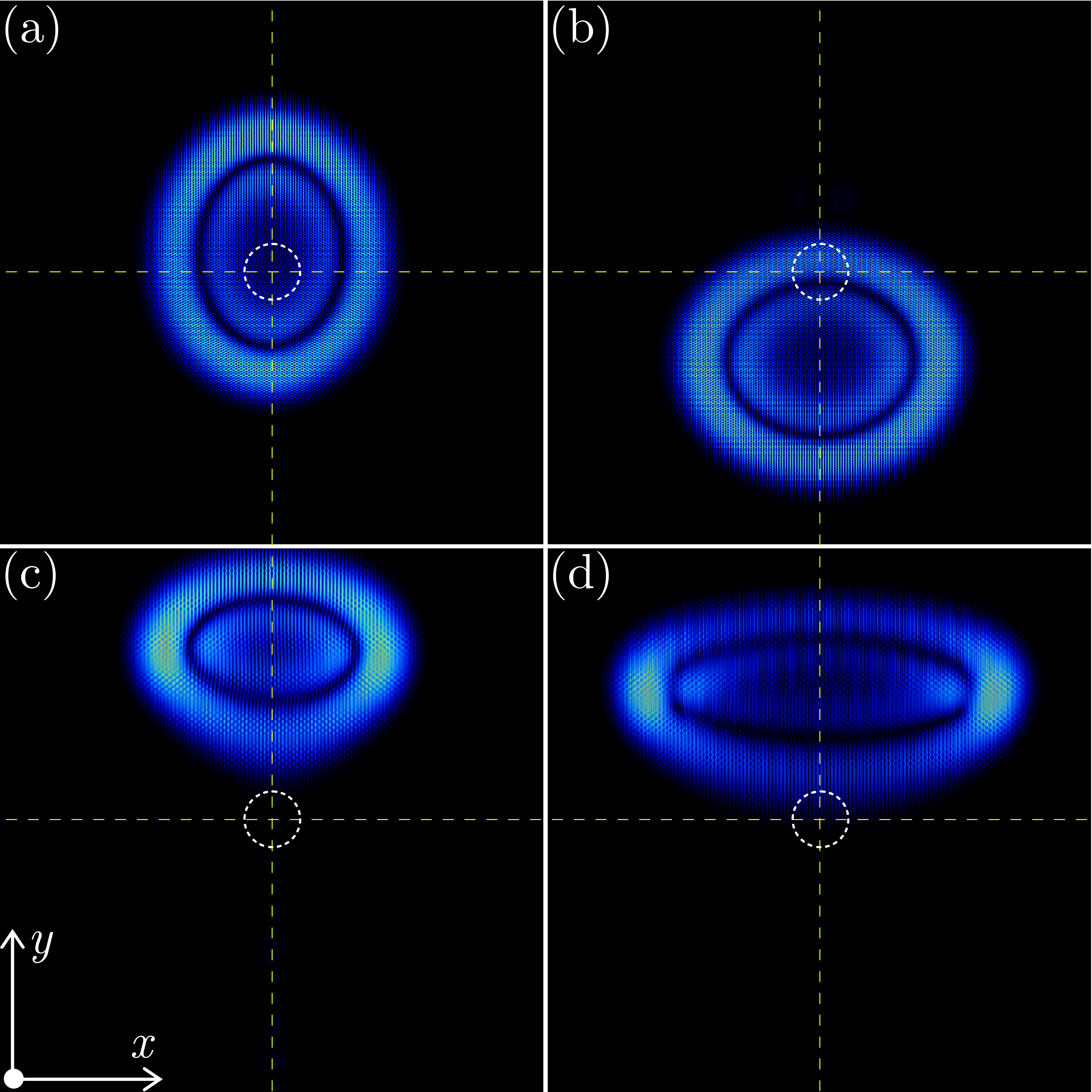}
  \caption{(a) Conical diffraction due to the tilted type-I Dirac cone marked by the green circle in Figure \ref{fig2}(b).
  (b) Conical diffraction due to the type-III Dirac cone marked by the red circle in Figure \ref{fig2}(b).
  (c) Conical diffraction due to the type-II Dirac cone marked by the green circle in Figure \ref{fig2}(c).
  (d) Conical diffraction due to the type-II Dirac cone marked by the red circle in Figure \ref{fig2}(c).
  All panels are in the range $-300\le x,y\le 300$.
  Dashed lines are along $x=0$ and $y=0$.
  Dotted circles in each panel represent the input beams.}
  \label{fig4}
\end{figure}

\section{Results}
\subsection{Conical diffraction}
If the incident beam can excite the Dirac cone states properly,
the beam will undergo conical diffraction during propagation
\cite{peleg.prl.98.103901.2007,ablowitz.pra.79.053830.2009,diebel.prl.116.183902.2016,leykam.aipx.1.101.2016,zhong.adp.529.1600258.2017,zhong.oe.27.6300.2019,kang.adp.531.1900295.2019}.
Since there are type-I, type-II and type-III Dirac cones in Figure \ref{fig2}(b) and \ref{fig2}(c),
we do not consider the Dirac cones in Figure \ref{fig2}(a) and only excite these Dirac cone states indicated by green and red circles,
which can be achieved via the method developed in Refs. \cite{zhong.oe.27.6300.2019,kang.adp.531.1900295.2019}.
After propagating a distance of $z=200$,
the output intensity profiles of these excited Dirac cone states are displayed in Figure \ref{fig4},
in which the dotted circles represent the location and size of the input Dirac cone states,
and the vertical and horizontal dashed yellow lines are the $x=0$ and $y=0$ axes, respectively.

Corresponding to the tilted type-I Dirac cone surrounded by the green circle in Figure \ref{fig2}(b),
the conical diffraction is displayed in Figure \ref{fig4}(a).
Due to the tilt of the Dirac cone, the diffraction speed along $+y$ and $-y$ directions are different,
so the diffraction ring exhibits an elliptic profile.
If the type-III Dirac cone state is excited,
as indicated by the red circle in Figure \ref{fig2}(b),
the conical diffraction may only happen along $-y$ direction,
because the group velocity along $+y$ direction $-d\beta/dk_y \approx 0$.
The numerical result is shown in Figure \ref{fig4}(b),
and indeed, the diffraction ring only appears in the $-y$ region
with the location of the upper edge of the ring nearly being invariant.
Different from the type-I and type-III Dirac cones,
the sign of speed $-d\beta/dk_y$ of the type-II Dirac cone does not change along $y$ coordinate,
so the whole conical diffraction ring will move along either $+y$ or $-y$ direction.
As to the type-II Dirac cones marked with green and red circles in Figure \ref{fig2}(c),
conical diffraction rings will move along $+y$ direction because of $-d\beta/dk_y>0$,
and the corresponding numerical simulations are displayed in Figure \ref{fig4}(c) and \ref{fig4}(d).
Considering that all the Dirac cones are tilted along $k_y$ coordinate only,
the conical diffractions in Figure \ref{fig4} are symmetric about $x=0$.
Note that the diameter of the diffraction ring along $x$ coordinate in Figure \ref{fig4}(d)
is much bigger than that in Figure \ref{fig4}(c),
and the reason is that the absolute value of $-d\beta/dk_x$ of the Dirac cone surrounded by the green circle
is smaller than that of the Dirac cone surrounded by the red circle.
In short, the spatial conical diffraction patterns can be predicted from the profiles of the Dirac cones,
and \textit{vice versa}, the properties of the Dirac cones are manifested into the conical diffraction patterns.

\begin{figure}
  \includegraphics[width=0.5\columnwidth]{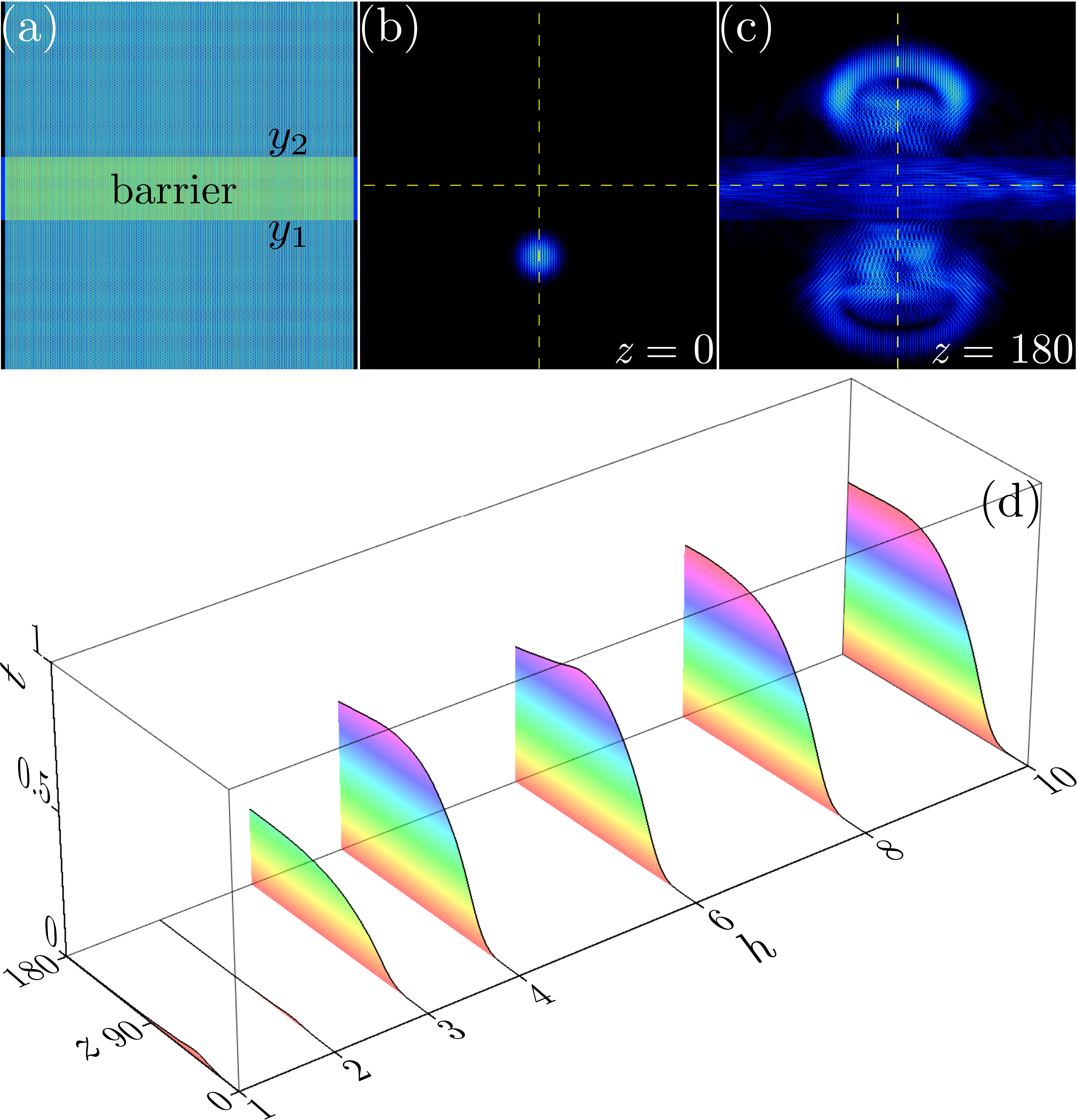}
  \caption{(a) Type-II photonic lattice with a barrier.
  (b) Input type-II Dirac state superimposed with a wide Gaussian, which is same as those used in Figure \ref{fig4} but with the beam center locating at $(0,-120)$.
  (c) Output of the beam which exhibits Klein tunneling with $h=10$, $y_1=-58$ and $y_2=48$.
  (d) Klein tunneling with different height of the barrier.
  Panels in (b,c) are in the range $-300\le x,y\le 300$, and dashed lines are along $x=0$ and $y=0$.}
  \label{fig5}
\end{figure}

\subsection{Klein tunneling}

Different from ordinary quantum mechanical tunneling,
the term Klein tunneling refers to a counterintuitive relativistic process
in which an electron can penetrate through a potential barrier
higher than the electron's rest energy \cite{katsnelson.np.2.620.2006}. 
Since conical diffraction due to the type-II Dirac cones moves along $+y$ direction spontaneously
even with zero incident angle,
it becomes an ideal paradigm to investigate the Klein tunneling.
This is indeed more advantage than that in type-I Dirac photonic lattices,
where the movement of the incident beam is required through tunning the angle of incidence.
What one is required to do is setting up a barrier in the type-II Dirac photonic lattice at a proper place
where the conical diffraction will move toward.
Generally, the lattice potential superimposed with a barrier can be written as
\begin{equation}\label{eq6}
  \mathcal{R}(x,y) =
  \left\{
  \begin{array}{ll}
\mathcal{R}(x,y) & \quad\textrm{if $y \le y_1$ or $y>y_2$},\\
\mathcal{R}(x,y)+h & \quad\textrm{if $y_1 < y \le y_2$},
\end{array}
  \right.
\end{equation}
in which $h>0$ is the height and $y_2-y_1$ is the width of the barrier.
The type-II Dirac photonic lattice with a barrier is shown in Figure \ref{fig5}(a),
and the incident type-II Dirac cone state [corresponding to the Dirac cone marked with the green circle in Figure \ref{fig2}(c)] superimposed with a wide Gaussian that is placed
below the barrier is displayed in Figure \ref{fig5}(b).
The incident state will exhibit conical diffraction during propagation,
and meanwhile, the conical diffraction pattern moves upward along $+y$ direction, as in Figure \ref{fig4}(c).
Inevitably, the conical diffraction pattern will hit on the barrier,
and it will be reflected by the barrier by intuition.
Nevertheless, transmission of the conical diffraction pattern over the barrier is completely possible
if the barrier height is bigger than 3 --  a requirement of Klein tunneling, because the ``energy'' of the state
at the Dirac point marked with the green circle in Figure \ref{fig2}(c) is about $\beta\approx3$.
To track the Klein tunneling process, we define a physical quantity named the transmission ratio,
as \[r=\frac{P_{\rm KT}}{P},\]
with \[P = \int_{-\infty}^{+\infty} \int_{-\infty}^{+\infty}  |\psi|^2dxdy\]
and \[P_{\rm KT} = \int_{-\infty}^{+\infty} \int_{y_1}^{+\infty}|\psi|^2 dx dy.\]
The transmission ratio $t$ as a function of propagation distance $z$ for barriers with different height
is shown in Figure \ref{fig5}(d).
Distinctly, the conical diffraction is almost reflected by the barrier
when the height is smaller than the ``energy'';
see the curves corresponding to $h=1$ and $h=2$ in Figure \ref{fig5}(d).
Klein tunneling starts to happen if the barrier height is close to the ``energy'',
and this is demonstrated by the curve with $h=3$.
Increasing the barrier height further, the transmission ratio also grows,
but numerical simulations indicate that there is seemingly a saturable value for the transmission ratio,
which is $\sim70\%$.
The reason for not reaching the perfect Klein tunneling is due to that
the incident beam in Figure \ref{fig5}(b) is not really ``massless''.
On one hand, the Dirac cone state is approximately obtained
and its corresponding Bloch momentum is not exactly in the
infinitesimal region around the critical point.
On the other hand, the superimposed wide Gaussian beam is a ``massive'' object.
The output amplitude profile with $h=10$ is displayed in Figure \ref{fig5}(c),
which clearly manifests the Klein tunneling of the conical diffraction.
Since the conical diffraction is still hold after penetrating the barrier,
excitation of the type-II Dirac cone mode is conserved.

Note that the steps of the barrier given by Equation (\ref{eq6}) are sharp.
If this barrier is replace by a barrier with smooth steps, e.g., a super-Gaussian,
which transforms the potential into
\[\mathcal{R}(x,y)=\mathcal{R}(x,y)+h\exp(-y^8/w^8)\] with $w=50$,
one will not miss out on the finding that the Klein tunneling is completely inhibited,
which also demonstrates that the phenomenon reported in Figure \ref{fig5} is truly Klein tunneling.
The explanation of this inhibition is that
the barrier with smooth steps can be regarded as a combination of sub-barriers with sharp steps, infinitesimal width,
and different height.
The conical diffraction will first encounter the sub-barrier with very small height upon its movement along $+y$ direction, and
this height is always smaller than the ``energy'' of the conical diffraction beam which dissatisfies the requirement of appearance of the Klein tunneling.

\section{Conclusion}
Summarizing, we have constructed a type-II Dirac photonic lattice from the Lieb-like lattice
by simply adjusting a angle parameter which only changes the spatial symmetry of the lattice.
Conical diffraction and Klein tunneling in this novel type-II Dirac photonic lattice are
discussed in detail.
We believe that our results may not only provide a feasible avenue on light manipulation,
but also help realize other topological photonic and analogize nonrelativistic phenomena in photonic lattices.
In addition to photonics, we believe that the developed type-II Dirac lattice in this work
may also provide a completely new flatform for cold atoms and acoustics.

Note added:
A paper \cite{wu.prl.124.075501.2020} on constructing type-II Dirac points by proposing a band-folding scheme,
and a paper \cite{pyrialakos.nc.11.2074.2020} on symmetry-controlled
edge states in the type-II phase of Dirac photonic lattices, were
published after submission of this paper.
We would like to emphasize that, in our work for the first time,
we find the simplest way to construct type-II Dirac points
which only depends on the spatial geometry of the photonic lattice.

\section*{Acknowledgements}
This work was supported
by Guangdong Basic and Applied Basic Research Foundation
(2018A0303130057), National Natural Science Foundation of China
(U1537210, 11534008), and Fundamental Research Funds for the Central
Universities (xzy012019038, xzy022019076). The authors acknowledge
the computational resources provided by the HPC platform of Xi¡¯an Jiaotong
University.

\bibliographystyle{myprx}
\bibliography{my_refs_library}

\begin{thebibliography}{60}%
\makeatletter
\providecommand \@ifxundefined [1]{%
 \@ifx{#1\undefined}
}%
\providecommand \@ifnum [1]{%
 \ifnum #1\expandafter \@firstoftwo
 \else \expandafter \@secondoftwo
 \fi
}%
\providecommand \@ifx [1]{%
 \ifx #1\expandafter \@firstoftwo
 \else \expandafter \@secondoftwo
 \fi
}%
\providecommand \natexlab [1]{#1}%
\providecommand \enquote  [1]{``#1''}%
\providecommand \bibnamefont  [1]{#1}%
\providecommand \bibfnamefont [1]{#1}%
\providecommand \citenamefont [1]{#1}%
\providecommand \href@noop [0]{\@secondoftwo}%
\providecommand \href [0]{\begingroup \@sanitize@url \@href}%
\providecommand \@href[1]{\@@startlink{#1}\@@href}%
\providecommand \@@href[1]{\endgroup#1\@@endlink}%
\providecommand \@sanitize@url [0]{\catcode `\\12\catcode `\$12\catcode
  `\&12\catcode `\#12\catcode `\^12\catcode `\_12\catcode `\%12\relax}%
\providecommand \@@startlink[1]{}%
\providecommand \@@endlink[0]{}%
\providecommand \url  [0]{\begingroup\@sanitize@url \@url }%
\providecommand \@url [1]{\endgroup\@href {#1}{\urlprefix }}%
\providecommand \urlprefix  [0]{URL }%
\providecommand \Eprint [0]{\href }%
\providecommand \doibase [0]{http://dx.doi.org/}%
\providecommand \selectlanguage [0]{\@gobble}%
\providecommand \bibinfo  [0]{\@secondoftwo}%
\providecommand \bibfield  [0]{\@secondoftwo}%
\providecommand \translation [1]{[#1]}%
\providecommand \BibitemOpen [0]{}%
\providecommand \bibitemStop [0]{}%
\providecommand \bibitemNoStop [0]{.\EOS\space}%
\providecommand \EOS [0]{\spacefactor3000\relax}%
\providecommand \BibitemShut  [1]{\csname bibitem#1\endcsname}%
\let\auto@bib@innerbib\@empty
\bibitem [{\citenamefont {Leykam}\ and\ \citenamefont
  {Desyatnikov}(2016)}]{leykam.aipx.1.101.2016}%
  \BibitemOpen
  \bibfield  {author} {\bibinfo {author} {\bibfnamefont {D.}~\bibnamefont
  {Leykam}}\ and\ \bibinfo {author} {\bibfnamefont {A.~S.}\ \bibnamefont
  {Desyatnikov}},\ }\emph {\bibinfo {title} {Conical intersections for light
  and matter waves}},\ \href {\doibase 10.1080/23746149.2016.1144482}
  {\bibfield  {journal} {\bibinfo  {journal} {Adv. Phys. X}\ }\textbf {\bibinfo
  {volume} {1}},\ \bibinfo {pages} {101} (\bibinfo {year} {2016})}\BibitemShut
  {NoStop}%
\bibitem [{\citenamefont {Zandbergen}\ and\ \citenamefont
  {de~Dood}(2010)}]{zandbergen.prl.104.043903.2010}%
  \BibitemOpen
  \bibfield  {author} {\bibinfo {author} {\bibfnamefont {S.~R.}\ \bibnamefont
  {Zandbergen}}\ and\ \bibinfo {author} {\bibfnamefont {M.~J.~A.}\ \bibnamefont
  {de~Dood}},\ }\emph {\bibinfo {title} {Experimental Observation of Strong
  Edge Effects on the Pseudodiffusive Transport of Light in Photonic
  Graphene}},\ \href {\doibase 10.1103/PhysRevLett.104.043903} {\bibfield
  {journal} {\bibinfo  {journal} {Phys. Rev. Lett.}\ }\textbf {\bibinfo
  {volume} {104}},\ \bibinfo {pages} {043903} (\bibinfo {year}
  {2010})}\BibitemShut {NoStop}%
\bibitem [{\citenamefont {Rechtsman}\ \emph
  {et~al.}(2013{\natexlab{a}})\citenamefont {Rechtsman}, \citenamefont
  {Plotnik}, \citenamefont {Zeuner}, \citenamefont {Song}, \citenamefont
  {Chen}, \citenamefont {Szameit},\ and\ \citenamefont
  {Segev}}]{rechtsman.prl.111.103901.2013}%
  \BibitemOpen
  \bibfield  {author} {\bibinfo {author} {\bibfnamefont {M.~C.}\ \bibnamefont
  {Rechtsman}}, \bibinfo {author} {\bibfnamefont {Y.}~\bibnamefont {Plotnik}},
  \bibinfo {author} {\bibfnamefont {J.~M.}\ \bibnamefont {Zeuner}}, \bibinfo
  {author} {\bibfnamefont {D.}~\bibnamefont {Song}}, \bibinfo {author}
  {\bibfnamefont {Z.}~\bibnamefont {Chen}}, \bibinfo {author} {\bibfnamefont
  {A.}~\bibnamefont {Szameit}}, \ and\ \bibinfo {author} {\bibfnamefont
  {M.}~\bibnamefont {Segev}},\ }\emph {\bibinfo {title} {Topological Creation
  and Destruction of Edge States in Photonic Graphene}},\ \href {\doibase
  10.1103/PhysRevLett.111.103901} {\bibfield  {journal} {\bibinfo  {journal}
  {Phys. Rev. Lett.}\ }\textbf {\bibinfo {volume} {111}},\ \bibinfo {pages}
  {103901} (\bibinfo {year} {2013}{\natexlab{a}})}\BibitemShut {NoStop}%
\bibitem [{\citenamefont {Rechtsman}\ \emph
  {et~al.}(2013{\natexlab{b}})\citenamefont {Rechtsman}, \citenamefont
  {Zeuner}, \citenamefont {T\"unnermann}, \citenamefont {Nolte}, \citenamefont
  {Segev},\ and\ \citenamefont {Szameit}}]{rechtsman.np.7.153.2013}%
  \BibitemOpen
  \bibfield  {author} {\bibinfo {author} {\bibfnamefont {M.~C.}\ \bibnamefont
  {Rechtsman}}, \bibinfo {author} {\bibfnamefont {J.~M.}\ \bibnamefont
  {Zeuner}}, \bibinfo {author} {\bibfnamefont {A.}~\bibnamefont
  {T\"unnermann}}, \bibinfo {author} {\bibfnamefont {S.}~\bibnamefont {Nolte}},
  \bibinfo {author} {\bibfnamefont {M.}~\bibnamefont {Segev}}, \ and\ \bibinfo
  {author} {\bibfnamefont {A.}~\bibnamefont {Szameit}},\ }\emph {\bibinfo
  {title} {Strain-induced pseudomagnetic field and photonic {Landau} levels in
  dielectric structures}},\ \href {\doibase 10.1038/nphoton.2012.302}
  {\bibfield  {journal} {\bibinfo  {journal} {Nat. Photon.}\ }\textbf {\bibinfo
  {volume} {7}},\ \bibinfo {pages} {153} (\bibinfo {year}
  {2013}{\natexlab{b}})}\BibitemShut {NoStop}%
\bibitem [{\citenamefont {Crespi}\ \emph {et~al.}(2013)\citenamefont {Crespi},
  \citenamefont {Corrielli}, \citenamefont {Valle}, \citenamefont {Osellame},\
  and\ \citenamefont {Longhi}}]{crespi.njp.15.013012.2013}%
  \BibitemOpen
  \bibfield  {author} {\bibinfo {author} {\bibfnamefont {A.}~\bibnamefont
  {Crespi}}, \bibinfo {author} {\bibfnamefont {G.}~\bibnamefont {Corrielli}},
  \bibinfo {author} {\bibfnamefont {G.~D.}\ \bibnamefont {Valle}}, \bibinfo
  {author} {\bibfnamefont {R.}~\bibnamefont {Osellame}}, \ and\ \bibinfo
  {author} {\bibfnamefont {S.}~\bibnamefont {Longhi}},\ }\emph {\bibinfo
  {title} {Dynamic band collapse in photonic graphene}},\ \href {\doibase
  10.1088/1367-2630/15/1/013012} {\bibfield  {journal} {\bibinfo  {journal}
  {New J. Phys.}\ }\textbf {\bibinfo {volume} {15}},\ \bibinfo {pages} {013012}
  (\bibinfo {year} {2013})}\BibitemShut {NoStop}%
\bibitem [{\citenamefont {Zeuner}\ \emph {et~al.}(2014)\citenamefont {Zeuner},
  \citenamefont {Rechtsman}, \citenamefont {Nolte},\ and\ \citenamefont
  {Szameit}}]{zeuner.ol.39.602.2014}%
  \BibitemOpen
  \bibfield  {author} {\bibinfo {author} {\bibfnamefont {J.~M.}\ \bibnamefont
  {Zeuner}}, \bibinfo {author} {\bibfnamefont {M.~C.}\ \bibnamefont
  {Rechtsman}}, \bibinfo {author} {\bibfnamefont {S.}~\bibnamefont {Nolte}}, \
  and\ \bibinfo {author} {\bibfnamefont {A.}~\bibnamefont {Szameit}},\ }\emph
  {\bibinfo {title} {Edge states in disordered photonic graphene}},\ \href
  {\doibase 10.1364/OL.39.000602} {\bibfield  {journal} {\bibinfo  {journal}
  {Opt. Lett.}\ }\textbf {\bibinfo {volume} {39}},\ \bibinfo {pages} {602}
  (\bibinfo {year} {2014})}\BibitemShut {NoStop}%
\bibitem [{\citenamefont {Plotnik}\ \emph {et~al.}(2014)\citenamefont
  {Plotnik}, \citenamefont {Rechtsman}, \citenamefont {Song}, \citenamefont
  {Heinrich}, \citenamefont {Zeuner}, \citenamefont {Nolte}, \citenamefont
  {Lumer}, \citenamefont {Malkova}, \citenamefont {Xu}, \citenamefont
  {Szameit}, \citenamefont {Chen},\ and\ \citenamefont
  {Segev}}]{plotnik.nm.13.57.2014}%
  \BibitemOpen
  \bibfield  {author} {\bibinfo {author} {\bibfnamefont {Y.}~\bibnamefont
  {Plotnik}}, \bibinfo {author} {\bibfnamefont {M.~C.}\ \bibnamefont
  {Rechtsman}}, \bibinfo {author} {\bibfnamefont {D.}~\bibnamefont {Song}},
  \bibinfo {author} {\bibfnamefont {M.}~\bibnamefont {Heinrich}}, \bibinfo
  {author} {\bibfnamefont {J.~M.}\ \bibnamefont {Zeuner}}, \bibinfo {author}
  {\bibfnamefont {S.}~\bibnamefont {Nolte}}, \bibinfo {author} {\bibfnamefont
  {Y.}~\bibnamefont {Lumer}}, \bibinfo {author} {\bibfnamefont
  {N.}~\bibnamefont {Malkova}}, \bibinfo {author} {\bibfnamefont
  {J.}~\bibnamefont {Xu}}, \bibinfo {author} {\bibfnamefont {A.}~\bibnamefont
  {Szameit}}, \bibinfo {author} {\bibfnamefont {Z.}~\bibnamefont {Chen}}, \
  and\ \bibinfo {author} {\bibfnamefont {M.}~\bibnamefont {Segev}},\ }\emph
  {\bibinfo {title} {Observation of unconventional edge states in `photonic
  graphene'}},\ \href {\doibase 10.1038/nmat3783} {\bibfield  {journal}
  {\bibinfo  {journal} {Nat. Mater.}\ }\textbf {\bibinfo {volume} {13}},\
  \bibinfo {pages} {57} (\bibinfo {year} {2014})}\BibitemShut {NoStop}%
\bibitem [{\citenamefont {Song}\ \emph {et~al.}(2015)\citenamefont {Song},
  \citenamefont {Paltoglou}, \citenamefont {Liu}, \citenamefont {Zhu},
  \citenamefont {Gallardo}, \citenamefont {Tang}, \citenamefont {Xu},
  \citenamefont {Ablowitz}, \citenamefont {Efremidis},\ and\ \citenamefont
  {Chen}}]{song.nc.6.6272.2015}%
  \BibitemOpen
  \bibfield  {author} {\bibinfo {author} {\bibfnamefont {D.}~\bibnamefont
  {Song}}, \bibinfo {author} {\bibfnamefont {V.}~\bibnamefont {Paltoglou}},
  \bibinfo {author} {\bibfnamefont {S.}~\bibnamefont {Liu}}, \bibinfo {author}
  {\bibfnamefont {Y.}~\bibnamefont {Zhu}}, \bibinfo {author} {\bibfnamefont
  {D.}~\bibnamefont {Gallardo}}, \bibinfo {author} {\bibfnamefont
  {L.}~\bibnamefont {Tang}}, \bibinfo {author} {\bibfnamefont {J.}~\bibnamefont
  {Xu}}, \bibinfo {author} {\bibfnamefont {M.}~\bibnamefont {Ablowitz}},
  \bibinfo {author} {\bibfnamefont {N.~K.}\ \bibnamefont {Efremidis}}, \ and\
  \bibinfo {author} {\bibfnamefont {Z.}~\bibnamefont {Chen}},\ }\emph {\bibinfo
  {title} {Unveiling pseudospin and angular momentum in photonic graphene.}},\
  \href {\doibase 10.1038/ncomms7272} {\bibfield  {journal} {\bibinfo
  {journal} {Nat. Commun.}\ }\textbf {\bibinfo {volume} {6}},\ \bibinfo {pages}
  {6272} (\bibinfo {year} {2015})}\BibitemShut {NoStop}%
\bibitem [{\citenamefont {Nalitov}\ \emph {et~al.}(2015)\citenamefont
  {Nalitov}, \citenamefont {Malpuech}, \citenamefont {Ter\c{c}as},\ and\
  \citenamefont {Solnyshkov}}]{nalitov.prl.114.026803.2015}%
  \BibitemOpen
  \bibfield  {author} {\bibinfo {author} {\bibfnamefont {A.~V.}\ \bibnamefont
  {Nalitov}}, \bibinfo {author} {\bibfnamefont {G.}~\bibnamefont {Malpuech}},
  \bibinfo {author} {\bibfnamefont {H.}~\bibnamefont {Ter\c{c}as}}, \ and\
  \bibinfo {author} {\bibfnamefont {D.~D.}\ \bibnamefont {Solnyshkov}},\ }\emph
  {\bibinfo {title} {Spin-Orbit Coupling and the Optical Spin Hall Effect in
  Photonic Graphene}},\ \href {\doibase 10.1103/PhysRevLett.114.026803}
  {\bibfield  {journal} {\bibinfo  {journal} {Phys. Rev. Lett.}\ }\textbf
  {\bibinfo {volume} {114}},\ \bibinfo {pages} {026803} (\bibinfo {year}
  {2015})}\BibitemShut {NoStop}%
\bibitem [{\citenamefont {Taie}\ \emph {et~al.}(2015)\citenamefont {Taie},
  \citenamefont {Ozawa}, \citenamefont {Ichinose}, \citenamefont {Nishio},
  \citenamefont {Nakajima},\ and\ \citenamefont
  {Takahashi}}]{taie.sa.1.1500854.2015}%
  \BibitemOpen
  \bibfield  {author} {\bibinfo {author} {\bibfnamefont {S.}~\bibnamefont
  {Taie}}, \bibinfo {author} {\bibfnamefont {H.}~\bibnamefont {Ozawa}},
  \bibinfo {author} {\bibfnamefont {T.}~\bibnamefont {Ichinose}}, \bibinfo
  {author} {\bibfnamefont {T.}~\bibnamefont {Nishio}}, \bibinfo {author}
  {\bibfnamefont {S.}~\bibnamefont {Nakajima}}, \ and\ \bibinfo {author}
  {\bibfnamefont {Y.}~\bibnamefont {Takahashi}},\ }\emph {\bibinfo {title}
  {Coherent driving and freezing of bosonic matter wave in an optical Lieb
  lattice}},\ \href {\doibase 10.1126/sciadv.1500854} {\bibfield  {journal}
  {\bibinfo  {journal} {Sci. Adv.}\ }\textbf {\bibinfo {volume} {1}},\ \bibinfo
  {pages} {1500854} (\bibinfo {year} {2015})}\BibitemShut {NoStop}%
\bibitem [{\citenamefont {Vicencio}\ \emph {et~al.}(2015)\citenamefont
  {Vicencio}, \citenamefont {Cantillano}, \citenamefont {Morales-Inostroza},
  \citenamefont {Real}, \citenamefont {Mej\'{i}a-Cort\'es}, \citenamefont
  {Weimann}, \citenamefont {Szameit},\ and\ \citenamefont
  {Molina}}]{vicencio.prl.114.245503.2015}%
  \BibitemOpen
  \bibfield  {author} {\bibinfo {author} {\bibfnamefont {R.~A.}\ \bibnamefont
  {Vicencio}}, \bibinfo {author} {\bibfnamefont {C.}~\bibnamefont
  {Cantillano}}, \bibinfo {author} {\bibfnamefont {L.}~\bibnamefont
  {Morales-Inostroza}}, \bibinfo {author} {\bibfnamefont {B.}~\bibnamefont
  {Real}}, \bibinfo {author} {\bibfnamefont {C.}~\bibnamefont
  {Mej\'{i}a-Cort\'es}}, \bibinfo {author} {\bibfnamefont {S.}~\bibnamefont
  {Weimann}}, \bibinfo {author} {\bibfnamefont {A.}~\bibnamefont {Szameit}}, \
  and\ \bibinfo {author} {\bibfnamefont {M.~I.}\ \bibnamefont {Molina}},\
  }\emph {\bibinfo {title} {Observation of Localized States in {L}ieb Photonic
  Lattices}},\ \href {\doibase 10.1103/PhysRevLett.114.245503} {\bibfield
  {journal} {\bibinfo  {journal} {Phys. Rev. Lett.}\ }\textbf {\bibinfo
  {volume} {114}},\ \bibinfo {pages} {245503} (\bibinfo {year}
  {2015})}\BibitemShut {NoStop}%
\bibitem [{\citenamefont {Mukherjee}\ \emph {et~al.}(2015)\citenamefont
  {Mukherjee}, \citenamefont {Spracklen}, \citenamefont {Choudhury},
  \citenamefont {Goldman}, \citenamefont {\"Ohberg}, \citenamefont
  {Andersson},\ and\ \citenamefont {Thomson}}]{mukherjee.prl.114.245504.2015}%
  \BibitemOpen
  \bibfield  {author} {\bibinfo {author} {\bibfnamefont {S.}~\bibnamefont
  {Mukherjee}}, \bibinfo {author} {\bibfnamefont {A.}~\bibnamefont
  {Spracklen}}, \bibinfo {author} {\bibfnamefont {D.}~\bibnamefont
  {Choudhury}}, \bibinfo {author} {\bibfnamefont {N.}~\bibnamefont {Goldman}},
  \bibinfo {author} {\bibfnamefont {P.}~\bibnamefont {\"Ohberg}}, \bibinfo
  {author} {\bibfnamefont {E.}~\bibnamefont {Andersson}}, \ and\ \bibinfo
  {author} {\bibfnamefont {R.~R.}\ \bibnamefont {Thomson}},\ }\emph {\bibinfo
  {title} {Observation of a Localized Flat-Band State in a Photonic {L}ieb
  Lattice}},\ \href {\doibase 10.1103/PhysRevLett.114.245504} {\bibfield
  {journal} {\bibinfo  {journal} {Phys. Rev. Lett.}\ }\textbf {\bibinfo
  {volume} {114}},\ \bibinfo {pages} {245504} (\bibinfo {year}
  {2015})}\BibitemShut {NoStop}%
\bibitem [{\citenamefont {Diebel}\ \emph {et~al.}(2016)\citenamefont {Diebel},
  \citenamefont {Leykam}, \citenamefont {Kroesen}, \citenamefont {Denz},\ and\
  \citenamefont {Desyatnikov}}]{diebel.prl.116.183902.2016}%
  \BibitemOpen
  \bibfield  {author} {\bibinfo {author} {\bibfnamefont {F.}~\bibnamefont
  {Diebel}}, \bibinfo {author} {\bibfnamefont {D.}~\bibnamefont {Leykam}},
  \bibinfo {author} {\bibfnamefont {S.}~\bibnamefont {Kroesen}}, \bibinfo
  {author} {\bibfnamefont {C.}~\bibnamefont {Denz}}, \ and\ \bibinfo {author}
  {\bibfnamefont {A.~S.}\ \bibnamefont {Desyatnikov}},\ }\emph {\bibinfo
  {title} {Conical Diffraction and Composite {L}ieb Bosons in Photonic
  Lattices}},\ \href {\doibase 10.1103/PhysRevLett.116.183902} {\bibfield
  {journal} {\bibinfo  {journal} {Phys. Rev. Lett.}\ }\textbf {\bibinfo
  {volume} {116}},\ \bibinfo {pages} {183902} (\bibinfo {year}
  {2016})}\BibitemShut {NoStop}%
\bibitem [{\citenamefont {Ozawa}\ \emph {et~al.}(2017)\citenamefont {Ozawa},
  \citenamefont {Taie}, \citenamefont {Ichinose},\ and\ \citenamefont
  {Takahashi}}]{ozawa.prl.118.175301.2017}%
  \BibitemOpen
  \bibfield  {author} {\bibinfo {author} {\bibfnamefont {H.}~\bibnamefont
  {Ozawa}}, \bibinfo {author} {\bibfnamefont {S.}~\bibnamefont {Taie}},
  \bibinfo {author} {\bibfnamefont {T.}~\bibnamefont {Ichinose}}, \ and\
  \bibinfo {author} {\bibfnamefont {Y.}~\bibnamefont {Takahashi}},\ }\emph
  {\bibinfo {title} {Interaction-Driven Shift and Distortion of a Flat Band in
  an Optical {Lieb} Lattice}},\ \href {\doibase 10.1103/PhysRevLett.118.175301}
  {\bibfield  {journal} {\bibinfo  {journal} {Phys. Rev. Lett.}\ }\textbf
  {\bibinfo {volume} {118}},\ \bibinfo {pages} {175301} (\bibinfo {year}
  {2017})}\BibitemShut {NoStop}%
\bibitem [{\citenamefont {Xia}\ \emph {et~al.}(2018)\citenamefont {Xia},
  \citenamefont {Ramachandran}, \citenamefont {Xia}, \citenamefont {Li},
  \citenamefont {Liu}, \citenamefont {Tang}, \citenamefont {Hu}, \citenamefont
  {Song}, \citenamefont {Xu}, \citenamefont {Leykam}, \citenamefont {Flach},\
  and\ \citenamefont {Chen}}]{xia.prl.121.263902.2018}%
  \BibitemOpen
  \bibfield  {author} {\bibinfo {author} {\bibfnamefont {S.}~\bibnamefont
  {Xia}}, \bibinfo {author} {\bibfnamefont {A.}~\bibnamefont {Ramachandran}},
  \bibinfo {author} {\bibfnamefont {S.}~\bibnamefont {Xia}}, \bibinfo {author}
  {\bibfnamefont {D.}~\bibnamefont {Li}}, \bibinfo {author} {\bibfnamefont
  {X.}~\bibnamefont {Liu}}, \bibinfo {author} {\bibfnamefont {L.}~\bibnamefont
  {Tang}}, \bibinfo {author} {\bibfnamefont {Y.}~\bibnamefont {Hu}}, \bibinfo
  {author} {\bibfnamefont {D.}~\bibnamefont {Song}}, \bibinfo {author}
  {\bibfnamefont {J.}~\bibnamefont {Xu}}, \bibinfo {author} {\bibfnamefont
  {D.}~\bibnamefont {Leykam}}, \bibinfo {author} {\bibfnamefont
  {S.}~\bibnamefont {Flach}}, \ and\ \bibinfo {author} {\bibfnamefont
  {Z.}~\bibnamefont {Chen}},\ }\emph {\bibinfo {title} {Unconventional Flatband
  Line States in Photonic {Lieb} Lattices}},\ \href {\doibase
  10.1103/PhysRevLett.121.263902} {\bibfield  {journal} {\bibinfo  {journal}
  {Phys. Rev. Lett.}\ }\textbf {\bibinfo {volume} {121}},\ \bibinfo {pages}
  {263902} (\bibinfo {year} {2018})}\BibitemShut {NoStop}%
\bibitem [{\citenamefont {El~Hassan}\ \emph {et~al.}(2019)\citenamefont
  {El~Hassan}, \citenamefont {Kunst}, \citenamefont {Moritz}, \citenamefont
  {Andler}, \citenamefont {Bergholtz},\ and\ \citenamefont
  {Bourennane}}]{hassan.np.13.697.2019}%
  \BibitemOpen
  \bibfield  {author} {\bibinfo {author} {\bibfnamefont {A.}~\bibnamefont
  {El~Hassan}}, \bibinfo {author} {\bibfnamefont {F.~K.}\ \bibnamefont
  {Kunst}}, \bibinfo {author} {\bibfnamefont {A.}~\bibnamefont {Moritz}},
  \bibinfo {author} {\bibfnamefont {G.}~\bibnamefont {Andler}}, \bibinfo
  {author} {\bibfnamefont {E.~J.}\ \bibnamefont {Bergholtz}}, \ and\ \bibinfo
  {author} {\bibfnamefont {M.}~\bibnamefont {Bourennane}},\ }\emph {\bibinfo
  {title} {Corner states of light in photonic waveguides}},\ \href
  {https://doi.org/10.1038/s41566-019-0519-y} {\bibfield  {journal} {\bibinfo
  {journal} {Nat. Photon.}\ }\textbf {\bibinfo {volume} {13}},\ \bibinfo
  {pages} {697} (\bibinfo {year} {2019})}\BibitemShut {NoStop}%
\bibitem [{\citenamefont {Li}\ \emph {et~al.}(2019{\natexlab{a}})\citenamefont
  {Li}, \citenamefont {Zhirihin}, \citenamefont {Gorlach}, \citenamefont {Ni},
  \citenamefont {Filonov}, \citenamefont {Slobozhanyuk}, \citenamefont {Alù},\
  and\ \citenamefont {Khanikaev}}]{li.np.2019}%
  \BibitemOpen
  \bibfield  {author} {\bibinfo {author} {\bibfnamefont {M.}~\bibnamefont
  {Li}}, \bibinfo {author} {\bibfnamefont {D.}~\bibnamefont {Zhirihin}},
  \bibinfo {author} {\bibfnamefont {M.}~\bibnamefont {Gorlach}}, \bibinfo
  {author} {\bibfnamefont {X.}~\bibnamefont {Ni}}, \bibinfo {author}
  {\bibfnamefont {D.}~\bibnamefont {Filonov}}, \bibinfo {author} {\bibfnamefont
  {A.}~\bibnamefont {Slobozhanyuk}}, \bibinfo {author} {\bibfnamefont
  {A.}~\bibnamefont {Alù}}, \ and\ \bibinfo {author} {\bibfnamefont {A.~B.}\
  \bibnamefont {Khanikaev}},\ }\emph {\bibinfo {title} {Higher-order
  topological states in photonic kagome crystals with long-range
  interactions}},\ \href {\doibase 10.1038/s41566-019-0561-9} {\bibfield
  {journal} {\bibinfo  {journal} {Nat. Photon.}\ } (\bibinfo {year}
  {2019}{\natexlab{a}}),\ 10.1038/s41566-019-0561-9}\BibitemShut {NoStop}%
\bibitem [{\citenamefont {Zhong}\ \emph {et~al.}(2017)\citenamefont {Zhong},
  \citenamefont {Zhang}, \citenamefont {Zhu}, \citenamefont {Zhang},
  \citenamefont {Li}, \citenamefont {Zhang}, \citenamefont {Li}, \citenamefont
  {Beli\'c},\ and\ \citenamefont {Xiao}}]{zhong.adp.529.1600258.2017}%
  \BibitemOpen
  \bibfield  {author} {\bibinfo {author} {\bibfnamefont {H.}~\bibnamefont
  {Zhong}}, \bibinfo {author} {\bibfnamefont {Y.~Q.}\ \bibnamefont {Zhang}},
  \bibinfo {author} {\bibfnamefont {Y.}~\bibnamefont {Zhu}}, \bibinfo {author}
  {\bibfnamefont {D.}~\bibnamefont {Zhang}}, \bibinfo {author} {\bibfnamefont
  {C.~B.}\ \bibnamefont {Li}}, \bibinfo {author} {\bibfnamefont {Y.~P.}\
  \bibnamefont {Zhang}}, \bibinfo {author} {\bibfnamefont {F.~L.}\ \bibnamefont
  {Li}}, \bibinfo {author} {\bibfnamefont {M.~R.}\ \bibnamefont {Beli\'c}}, \
  and\ \bibinfo {author} {\bibfnamefont {M.}~\bibnamefont {Xiao}},\ }\emph
  {\bibinfo {title} {Transport properties in the photonic super-honeycomb
  lattice -- a hybrid fermionic and bosonic system}},\ \href {\doibase
  10.1002/andp.201600258} {\bibfield  {journal} {\bibinfo  {journal} {Ann.
  Phys. (Berlin)}\ }\textbf {\bibinfo {volume} {529}},\ \bibinfo {pages}
  {1600258} (\bibinfo {year} {2017})}\BibitemShut {NoStop}%
\bibitem [{\citenamefont {Kang}\ \emph {et~al.}(2019)\citenamefont {Kang},
  \citenamefont {Zhong}, \citenamefont {Beli\'c}, \citenamefont {Tian},
  \citenamefont {Jin}, \citenamefont {Zhang}, \citenamefont {Li},\ and\
  \citenamefont {Zhang}}]{kang.adp.531.1900295.2019}%
  \BibitemOpen
  \bibfield  {author} {\bibinfo {author} {\bibfnamefont {Y.~F.}\ \bibnamefont
  {Kang}}, \bibinfo {author} {\bibfnamefont {H.}~\bibnamefont {Zhong}},
  \bibinfo {author} {\bibfnamefont {M.~R.}\ \bibnamefont {Beli\'c}}, \bibinfo
  {author} {\bibfnamefont {Y.~Q.}\ \bibnamefont {Tian}}, \bibinfo {author}
  {\bibfnamefont {K.~C.}\ \bibnamefont {Jin}}, \bibinfo {author} {\bibfnamefont
  {Y.~P.}\ \bibnamefont {Zhang}}, \bibinfo {author} {\bibfnamefont {F.~L.}\
  \bibnamefont {Li}}, \ and\ \bibinfo {author} {\bibfnamefont {Y.~Q.}\
  \bibnamefont {Zhang}},\ }\emph {\bibinfo {title} {Conical Diffraction from
  Approximate {D}irac Cone States in a Superhoneycomb Lattice}},\ \href
  {\doibase 10.1002/andp.201900295} {\bibfield  {journal} {\bibinfo  {journal}
  {Ann. Phys. (Berlin)}\ }\textbf {\bibinfo {volume} {531}},\ \bibinfo {pages}
  {1900295} (\bibinfo {year} {2019})}\BibitemShut {NoStop}%
\bibitem [{\citenamefont {Zhong}\ \emph {et~al.}(2019)\citenamefont {Zhong},
  \citenamefont {Wang}, \citenamefont {Beli\'{c}}, \citenamefont {Zhang},\ and\
  \citenamefont {Zhang}}]{zhong.oe.27.6300.2019}%
  \BibitemOpen
  \bibfield  {author} {\bibinfo {author} {\bibfnamefont {H.}~\bibnamefont
  {Zhong}}, \bibinfo {author} {\bibfnamefont {R.}~\bibnamefont {Wang}},
  \bibinfo {author} {\bibfnamefont {M.~R.}\ \bibnamefont {Beli\'{c}}}, \bibinfo
  {author} {\bibfnamefont {Y.~P.}\ \bibnamefont {Zhang}}, \ and\ \bibinfo
  {author} {\bibfnamefont {Y.~Q.}\ \bibnamefont {Zhang}},\ }\emph {\bibinfo
  {title} {Asymmetric conical diffraction in dislocated edge-centered square
  lattices}},\ \href {\doibase 10.1364/OE.27.006300} {\bibfield  {journal}
  {\bibinfo  {journal} {Opt. Express}\ }\textbf {\bibinfo {volume} {27}},\
  \bibinfo {pages} {6300} (\bibinfo {year} {2019})}\BibitemShut {NoStop}%
\bibitem [{\citenamefont {Mili\'{c}evi\'{c}}\ \emph {et~al.}(2019)\citenamefont
  {Mili\'{c}evi\'{c}}, \citenamefont {Montambaux}, \citenamefont {Ozawa},
  \citenamefont {Jamadi}, \citenamefont {Real}, \citenamefont {Sagnes},
  \citenamefont {Lema\^{\i}tre}, \citenamefont {Le~Gratiet}, \citenamefont
  {Harouri}, \citenamefont {Bloch},\ and\ \citenamefont
  {Amo}}]{milicevic.prx.9.031010.2019}%
  \BibitemOpen
  \bibfield  {author} {\bibinfo {author} {\bibfnamefont {M.}~\bibnamefont
  {Mili\'{c}evi\'{c}}}, \bibinfo {author} {\bibfnamefont {G.}~\bibnamefont
  {Montambaux}}, \bibinfo {author} {\bibfnamefont {T.}~\bibnamefont {Ozawa}},
  \bibinfo {author} {\bibfnamefont {O.}~\bibnamefont {Jamadi}}, \bibinfo
  {author} {\bibfnamefont {B.}~\bibnamefont {Real}}, \bibinfo {author}
  {\bibfnamefont {I.}~\bibnamefont {Sagnes}}, \bibinfo {author} {\bibfnamefont
  {A.}~\bibnamefont {Lema\^{\i}tre}}, \bibinfo {author} {\bibfnamefont
  {L.}~\bibnamefont {Le~Gratiet}}, \bibinfo {author} {\bibfnamefont
  {A.}~\bibnamefont {Harouri}}, \bibinfo {author} {\bibfnamefont
  {J.}~\bibnamefont {Bloch}}, \ and\ \bibinfo {author} {\bibfnamefont
  {A.}~\bibnamefont {Amo}},\ }\emph {\bibinfo {title} {Type-III and Tilted
  {D}irac Cones Emerging from Flat Bands in Photonic Orbital Graphene}},\ \href
  {\doibase 10.1103/PhysRevX.9.031010} {\bibfield  {journal} {\bibinfo
  {journal} {Phys. Rev. X}\ }\textbf {\bibinfo {volume} {9}},\ \bibinfo {pages}
  {031010} (\bibinfo {year} {2019})}\BibitemShut {NoStop}%
\bibitem [{\citenamefont {Li}\ \emph {et~al.}(2019{\natexlab{b}})\citenamefont
  {Li}, \citenamefont {Deng}, \citenamefont {Fu}, \citenamefont {Li},
  \citenamefont {Ma}, \citenamefont {Han}, \citenamefont {Zhou}, \citenamefont
  {Zhou},\ and\ \citenamefont {Yao}}]{li.arxiv}%
  \BibitemOpen
  \bibfield  {author} {\bibinfo {author} {\bibfnamefont {X.-P.}\ \bibnamefont
  {Li}}, \bibinfo {author} {\bibfnamefont {K.}~\bibnamefont {Deng}}, \bibinfo
  {author} {\bibfnamefont {B.}~\bibnamefont {Fu}}, \bibinfo {author}
  {\bibfnamefont {Y.}~\bibnamefont {Li}}, \bibinfo {author} {\bibfnamefont
  {D.}~\bibnamefont {Ma}}, \bibinfo {author} {\bibfnamefont {J.}~\bibnamefont
  {Han}}, \bibinfo {author} {\bibfnamefont {J.}~\bibnamefont {Zhou}}, \bibinfo
  {author} {\bibfnamefont {S.}~\bibnamefont {Zhou}}, \ and\ \bibinfo {author}
  {\bibfnamefont {Y.}~\bibnamefont {Yao}},\ }\emph {\bibinfo {title}
  {Type-{III} {W}eyl Semimetals and its Materialization}},\ \href@noop {}
  {\bibfield  {journal} {\bibinfo  {journal} {arXiv:1909.12178}\ } (\bibinfo
  {year} {2019}{\natexlab{b}})}\BibitemShut {NoStop}%
\bibitem [{\citenamefont {Soluyanov}\ \emph {et~al.}(2015)\citenamefont
  {Soluyanov}, \citenamefont {Gresch}, \citenamefont {Wang}, \citenamefont
  {Wu}, \citenamefont {Troyer}, \citenamefont {Dai},\ and\ \citenamefont
  {Bernevig}}]{soluyanov.nature.527.495.2015}%
  \BibitemOpen
  \bibfield  {author} {\bibinfo {author} {\bibfnamefont {A.~A.}\ \bibnamefont
  {Soluyanov}}, \bibinfo {author} {\bibfnamefont {D.}~\bibnamefont {Gresch}},
  \bibinfo {author} {\bibfnamefont {Z.}~\bibnamefont {Wang}}, \bibinfo {author}
  {\bibfnamefont {Q.}~\bibnamefont {Wu}}, \bibinfo {author} {\bibfnamefont
  {M.}~\bibnamefont {Troyer}}, \bibinfo {author} {\bibfnamefont
  {X.}~\bibnamefont {Dai}}, \ and\ \bibinfo {author} {\bibfnamefont {B.~A.}\
  \bibnamefont {Bernevig}},\ }\emph {\bibinfo {title} {Type-{II} {W}eyl
  semimetals}},\ \href {https://doi.org/10.1038/nature15768} {\bibfield
  {journal} {\bibinfo  {journal} {Nature}\ }\textbf {\bibinfo {volume} {527}},\
  \bibinfo {pages} {495} (\bibinfo {year} {2015})}\BibitemShut {NoStop}%
\bibitem [{\citenamefont {Xu}\ \emph {et~al.}(2015)\citenamefont {Xu},
  \citenamefont {Zhang},\ and\ \citenamefont {Zhang}}]{xu.prl.115.265304.2015}%
  \BibitemOpen
  \bibfield  {author} {\bibinfo {author} {\bibfnamefont {Y.}~\bibnamefont
  {Xu}}, \bibinfo {author} {\bibfnamefont {F.}~\bibnamefont {Zhang}}, \ and\
  \bibinfo {author} {\bibfnamefont {C.}~\bibnamefont {Zhang}},\ }\emph
  {\bibinfo {title} {Structured Weyl Points in Spin-Orbit Coupled Fermionic
  Superfluids}},\ \href {\doibase 10.1103/PhysRevLett.115.265304} {\bibfield
  {journal} {\bibinfo  {journal} {Phys. Rev. Lett.}\ }\textbf {\bibinfo
  {volume} {115}},\ \bibinfo {pages} {265304} (\bibinfo {year}
  {2015})}\BibitemShut {NoStop}%
\bibitem [{\citenamefont {Chen}\ \emph {et~al.}(2016)\citenamefont {Chen},
  \citenamefont {Xiao},\ and\ \citenamefont {Chan}}]{chen.nc.7.13038.2016}%
  \BibitemOpen
  \bibfield  {author} {\bibinfo {author} {\bibfnamefont {W.-J.}\ \bibnamefont
  {Chen}}, \bibinfo {author} {\bibfnamefont {M.}~\bibnamefont {Xiao}}, \ and\
  \bibinfo {author} {\bibfnamefont {C.~T.}\ \bibnamefont {Chan}},\ }\emph
  {\bibinfo {title} {Photonic crystals possessing multiple {W}eyl points and
  the experimental observation of robust surface states}},\ \href
  {https://doi.org/10.1038/ncomms13038} {\bibfield  {journal} {\bibinfo
  {journal} {Nat. Commun.}\ }\textbf {\bibinfo {volume} {7}},\ \bibinfo {pages}
  {13038} (\bibinfo {year} {2016})}\BibitemShut {NoStop}%
\bibitem [{\citenamefont {Aut\`es}\ \emph {et~al.}(2016)\citenamefont
  {Aut\`es}, \citenamefont {Gresch}, \citenamefont {Troyer}, \citenamefont
  {Soluyanov},\ and\ \citenamefont {Yazyev}}]{autes.prl.117.066402.2016}%
  \BibitemOpen
  \bibfield  {author} {\bibinfo {author} {\bibfnamefont {G.}~\bibnamefont
  {Aut\`es}}, \bibinfo {author} {\bibfnamefont {D.}~\bibnamefont {Gresch}},
  \bibinfo {author} {\bibfnamefont {M.}~\bibnamefont {Troyer}}, \bibinfo
  {author} {\bibfnamefont {A.~A.}\ \bibnamefont {Soluyanov}}, \ and\ \bibinfo
  {author} {\bibfnamefont {O.~V.}\ \bibnamefont {Yazyev}},\ }\emph {\bibinfo
  {title} {Robust Type-{II} {W}eyl Semimetal Phase in Transition Metal
  Diphosphides $X{\mathrm{P}}_{2}$ ($X=\mathrm{Mo}$, W)}},\ \href {\doibase
  10.1103/PhysRevLett.117.066402} {\bibfield  {journal} {\bibinfo  {journal}
  {Phys. Rev. Lett.}\ }\textbf {\bibinfo {volume} {117}},\ \bibinfo {pages}
  {066402} (\bibinfo {year} {2016})}\BibitemShut {NoStop}%
\bibitem [{\citenamefont {Noh}\ \emph {et~al.}(2017{\natexlab{a}})\citenamefont
  {Noh}, \citenamefont {Huang}, \citenamefont {Leykam}, \citenamefont {Chong},
  \citenamefont {Chen},\ and\ \citenamefont {Rechtsman}}]{noh.np.13.611.2017}%
  \BibitemOpen
  \bibfield  {author} {\bibinfo {author} {\bibfnamefont {J.}~\bibnamefont
  {Noh}}, \bibinfo {author} {\bibfnamefont {S.}~\bibnamefont {Huang}}, \bibinfo
  {author} {\bibfnamefont {D.}~\bibnamefont {Leykam}}, \bibinfo {author}
  {\bibfnamefont {Y.}~\bibnamefont {Chong}}, \bibinfo {author} {\bibfnamefont
  {K.~P.}\ \bibnamefont {Chen}}, \ and\ \bibinfo {author} {\bibfnamefont
  {M.}~\bibnamefont {Rechtsman}},\ }\emph {\bibinfo {title} {Experimental
  observation of optical {W}eyl points and {F}ermi arc-like surface states}},\
  \href {https://doi.org/10.1038/nphys4072} {\bibfield  {journal} {\bibinfo
  {journal} {Nat. Phys.}\ }\textbf {\bibinfo {volume} {13}},\ \bibinfo {pages}
  {611} (\bibinfo {year} {2017}{\natexlab{a}})}\BibitemShut {NoStop}%
\bibitem [{\citenamefont {Chen}\ \emph {et~al.}(2018)\citenamefont {Chen},
  \citenamefont {Zhou},\ and\ \citenamefont {Xu}}]{chen.prb.97.155152.2018}%
  \BibitemOpen
  \bibfield  {author} {\bibinfo {author} {\bibfnamefont {R.}~\bibnamefont
  {Chen}}, \bibinfo {author} {\bibfnamefont {B.}~\bibnamefont {Zhou}}, \ and\
  \bibinfo {author} {\bibfnamefont {D.-H.}\ \bibnamefont {Xu}},\ }\emph
  {\bibinfo {title} {Floquet {W}eyl semimetals in light-irradiated type-{II}
  and hybrid line-node semimetals}},\ \href {\doibase
  10.1103/PhysRevB.97.155152} {\bibfield  {journal} {\bibinfo  {journal} {Phys.
  Rev. B}\ }\textbf {\bibinfo {volume} {97}},\ \bibinfo {pages} {155152}
  (\bibinfo {year} {2018})}\BibitemShut {NoStop}%
\bibitem [{\citenamefont {Xie}\ \emph {et~al.}(2019)\citenamefont {Xie},
  \citenamefont {Liu}, \citenamefont {Cheng}, \citenamefont {Liu},
  \citenamefont {Chen},\ and\ \citenamefont {Tian}}]{xie.prl.122.104302.2019}%
  \BibitemOpen
  \bibfield  {author} {\bibinfo {author} {\bibfnamefont {B.}~\bibnamefont
  {Xie}}, \bibinfo {author} {\bibfnamefont {H.}~\bibnamefont {Liu}}, \bibinfo
  {author} {\bibfnamefont {H.}~\bibnamefont {Cheng}}, \bibinfo {author}
  {\bibfnamefont {Z.}~\bibnamefont {Liu}}, \bibinfo {author} {\bibfnamefont
  {S.}~\bibnamefont {Chen}}, \ and\ \bibinfo {author} {\bibfnamefont
  {J.}~\bibnamefont {Tian}},\ }\emph {\bibinfo {title} {Experimental
  Realization of Type-{II} {W}eyl Points and {F}ermi Arcs in Phononic
  Crystal}},\ \href {\doibase 10.1103/PhysRevLett.122.104302} {\bibfield
  {journal} {\bibinfo  {journal} {Phys. Rev. Lett.}\ }\textbf {\bibinfo
  {volume} {122}},\ \bibinfo {pages} {104302} (\bibinfo {year}
  {2019})}\BibitemShut {NoStop}%
\bibitem [{\citenamefont {Yan}\ \emph {et~al.}(2017)\citenamefont {Yan},
  \citenamefont {Huang}, \citenamefont {Zhang}, \citenamefont {Wang},
  \citenamefont {Yao}, \citenamefont {Deng}, \citenamefont {Wan}, \citenamefont
  {Zhang}, \citenamefont {Arita}, \citenamefont {Yang}, \citenamefont {Sun},
  \citenamefont {Yao}, \citenamefont {Wu}, \citenamefont {Fan}, \citenamefont
  {Duan},\ and\ \citenamefont {Zhou}}]{yan.nc.8.257.2017}%
  \BibitemOpen
  \bibfield  {author} {\bibinfo {author} {\bibfnamefont {M.}~\bibnamefont
  {Yan}}, \bibinfo {author} {\bibfnamefont {H.}~\bibnamefont {Huang}}, \bibinfo
  {author} {\bibfnamefont {K.}~\bibnamefont {Zhang}}, \bibinfo {author}
  {\bibfnamefont {E.}~\bibnamefont {Wang}}, \bibinfo {author} {\bibfnamefont
  {W.}~\bibnamefont {Yao}}, \bibinfo {author} {\bibfnamefont {K.}~\bibnamefont
  {Deng}}, \bibinfo {author} {\bibfnamefont {G.}~\bibnamefont {Wan}}, \bibinfo
  {author} {\bibfnamefont {H.}~\bibnamefont {Zhang}}, \bibinfo {author}
  {\bibfnamefont {M.}~\bibnamefont {Arita}}, \bibinfo {author} {\bibfnamefont
  {H.}~\bibnamefont {Yang}}, \bibinfo {author} {\bibfnamefont {Z.}~\bibnamefont
  {Sun}}, \bibinfo {author} {\bibfnamefont {H.}~\bibnamefont {Yao}}, \bibinfo
  {author} {\bibfnamefont {Y.}~\bibnamefont {Wu}}, \bibinfo {author}
  {\bibfnamefont {S.}~\bibnamefont {Fan}}, \bibinfo {author} {\bibfnamefont
  {W.}~\bibnamefont {Duan}}, \ and\ \bibinfo {author} {\bibfnamefont
  {S.}~\bibnamefont {Zhou}},\ }\emph {\bibinfo {title} {Lorentz-violating
  type-{II} {D}irac fermions in transition metal dichalcogenide PtTe$_2$}},\
  \href {https://doi.org/10.1038/s41467-017-00280-6} {\bibfield  {journal}
  {\bibinfo  {journal} {Nat. Commun.}\ }\textbf {\bibinfo {volume} {8}},\
  \bibinfo {pages} {257} (\bibinfo {year} {2017})}\BibitemShut {NoStop}%
\bibitem [{\citenamefont {Noh}\ \emph {et~al.}(2017{\natexlab{b}})\citenamefont
  {Noh}, \citenamefont {Jeong}, \citenamefont {Cho}, \citenamefont {Kim},
  \citenamefont {Min},\ and\ \citenamefont {Park}}]{noh.prl.119.016401.2017}%
  \BibitemOpen
  \bibfield  {author} {\bibinfo {author} {\bibfnamefont {H.-J.}\ \bibnamefont
  {Noh}}, \bibinfo {author} {\bibfnamefont {J.}~\bibnamefont {Jeong}}, \bibinfo
  {author} {\bibfnamefont {E.-J.}\ \bibnamefont {Cho}}, \bibinfo {author}
  {\bibfnamefont {K.}~\bibnamefont {Kim}}, \bibinfo {author} {\bibfnamefont
  {B.~I.}\ \bibnamefont {Min}}, \ and\ \bibinfo {author} {\bibfnamefont
  {B.-G.}\ \bibnamefont {Park}},\ }\emph {\bibinfo {title} {Experimental
  Realization of Type-{II Dirac} Fermions in a ${\mathrm{PdTe}}_{2}$
  Superconductor}},\ \href {\doibase 10.1103/PhysRevLett.119.016401} {\bibfield
   {journal} {\bibinfo  {journal} {Phys. Rev. Lett.}\ }\textbf {\bibinfo
  {volume} {119}},\ \bibinfo {pages} {016401} (\bibinfo {year}
  {2017}{\natexlab{b}})}\BibitemShut {NoStop}%
\bibitem [{\citenamefont {Chang}\ \emph {et~al.}(2017)\citenamefont {Chang},
  \citenamefont {Xu}, \citenamefont {Sanchez}, \citenamefont {Tsai},
  \citenamefont {Huang}, \citenamefont {Chang}, \citenamefont {Hsu},
  \citenamefont {Bian}, \citenamefont {Belopolski}, \citenamefont {Yu},
  \citenamefont {Yang}, \citenamefont {Neupert}, \citenamefont {Jeng},
  \citenamefont {Lin},\ and\ \citenamefont
  {Hasan}}]{chang.prl.119.026404.2017}%
  \BibitemOpen
  \bibfield  {author} {\bibinfo {author} {\bibfnamefont {T.-R.}\ \bibnamefont
  {Chang}}, \bibinfo {author} {\bibfnamefont {S.-Y.}\ \bibnamefont {Xu}},
  \bibinfo {author} {\bibfnamefont {D.~S.}\ \bibnamefont {Sanchez}}, \bibinfo
  {author} {\bibfnamefont {W.-F.}\ \bibnamefont {Tsai}}, \bibinfo {author}
  {\bibfnamefont {S.-M.}\ \bibnamefont {Huang}}, \bibinfo {author}
  {\bibfnamefont {G.}~\bibnamefont {Chang}}, \bibinfo {author} {\bibfnamefont
  {C.-H.}\ \bibnamefont {Hsu}}, \bibinfo {author} {\bibfnamefont
  {G.}~\bibnamefont {Bian}}, \bibinfo {author} {\bibfnamefont {I.}~\bibnamefont
  {Belopolski}}, \bibinfo {author} {\bibfnamefont {Z.-M.}\ \bibnamefont {Yu}},
  \bibinfo {author} {\bibfnamefont {S.~A.}\ \bibnamefont {Yang}}, \bibinfo
  {author} {\bibfnamefont {T.}~\bibnamefont {Neupert}}, \bibinfo {author}
  {\bibfnamefont {H.-T.}\ \bibnamefont {Jeng}}, \bibinfo {author}
  {\bibfnamefont {H.}~\bibnamefont {Lin}}, \ and\ \bibinfo {author}
  {\bibfnamefont {M.~Z.}\ \bibnamefont {Hasan}},\ }\emph {\bibinfo {title}
  {Type-{II} Symmetry-Protected Topological {D}irac Semimetals}},\ \href
  {\doibase 10.1103/PhysRevLett.119.026404} {\bibfield  {journal} {\bibinfo
  {journal} {Phys. Rev. Lett.}\ }\textbf {\bibinfo {volume} {119}},\ \bibinfo
  {pages} {026404} (\bibinfo {year} {2017})}\BibitemShut {NoStop}%
\bibitem [{\citenamefont {Fei}\ \emph {et~al.}(2018)\citenamefont {Fei},
  \citenamefont {Bo}, \citenamefont {Wang}, \citenamefont {Ying}, \citenamefont
  {Li}, \citenamefont {Chen}, \citenamefont {Dai}, \citenamefont {Chen},
  \citenamefont {Sun}, \citenamefont {Zhang}, \citenamefont {Qu}, \citenamefont
  {Zhang}, \citenamefont {Wang}, \citenamefont {Wang}, \citenamefont {Cao},
  \citenamefont {Bu}, \citenamefont {Song}, \citenamefont {Wan},\ and\
  \citenamefont {Wang}}]{fei.am.30.1801556.2018}%
  \BibitemOpen
  \bibfield  {author} {\bibinfo {author} {\bibfnamefont {F.}~\bibnamefont
  {Fei}}, \bibinfo {author} {\bibfnamefont {X.}~\bibnamefont {Bo}}, \bibinfo
  {author} {\bibfnamefont {P.}~\bibnamefont {Wang}}, \bibinfo {author}
  {\bibfnamefont {J.}~\bibnamefont {Ying}}, \bibinfo {author} {\bibfnamefont
  {J.}~\bibnamefont {Li}}, \bibinfo {author} {\bibfnamefont {K.}~\bibnamefont
  {Chen}}, \bibinfo {author} {\bibfnamefont {Q.}~\bibnamefont {Dai}}, \bibinfo
  {author} {\bibfnamefont {B.}~\bibnamefont {Chen}}, \bibinfo {author}
  {\bibfnamefont {Z.}~\bibnamefont {Sun}}, \bibinfo {author} {\bibfnamefont
  {M.}~\bibnamefont {Zhang}}, \bibinfo {author} {\bibfnamefont
  {F.}~\bibnamefont {Qu}}, \bibinfo {author} {\bibfnamefont {Y.}~\bibnamefont
  {Zhang}}, \bibinfo {author} {\bibfnamefont {Q.}~\bibnamefont {Wang}},
  \bibinfo {author} {\bibfnamefont {X.}~\bibnamefont {Wang}}, \bibinfo {author}
  {\bibfnamefont {L.}~\bibnamefont {Cao}}, \bibinfo {author} {\bibfnamefont
  {H.}~\bibnamefont {Bu}}, \bibinfo {author} {\bibfnamefont {F.}~\bibnamefont
  {Song}}, \bibinfo {author} {\bibfnamefont {X.}~\bibnamefont {Wan}}, \ and\
  \bibinfo {author} {\bibfnamefont {B.}~\bibnamefont {Wang}},\ }\emph {\bibinfo
  {title} {Band Structure Perfection and Superconductivity in Type-{II} Dirac
  Semimetal Ir$_{1-x}$Pt$_x$Te$_2$}},\ \href {\doibase 10.1002/adma.201801556}
  {\bibfield  {journal} {\bibinfo  {journal} {Adv. Mater.}\ }\textbf {\bibinfo
  {volume} {30}},\ \bibinfo {pages} {1801556} (\bibinfo {year}
  {2018})}\BibitemShut {NoStop}%
\bibitem [{\citenamefont {Politano}\ \emph {et~al.}(2018)\citenamefont
  {Politano}, \citenamefont {Chiarello}, \citenamefont {Ghosh}, \citenamefont
  {Sadhukhan}, \citenamefont {Kuo}, \citenamefont {Lue}, \citenamefont
  {Pellegrini},\ and\ \citenamefont {Agarwal}}]{politano.prl.121.086804.2018}%
  \BibitemOpen
  \bibfield  {author} {\bibinfo {author} {\bibfnamefont {A.}~\bibnamefont
  {Politano}}, \bibinfo {author} {\bibfnamefont {G.}~\bibnamefont {Chiarello}},
  \bibinfo {author} {\bibfnamefont {B.}~\bibnamefont {Ghosh}}, \bibinfo
  {author} {\bibfnamefont {K.}~\bibnamefont {Sadhukhan}}, \bibinfo {author}
  {\bibfnamefont {C.-N.}\ \bibnamefont {Kuo}}, \bibinfo {author} {\bibfnamefont
  {C.~S.}\ \bibnamefont {Lue}}, \bibinfo {author} {\bibfnamefont
  {V.}~\bibnamefont {Pellegrini}}, \ and\ \bibinfo {author} {\bibfnamefont
  {A.}~\bibnamefont {Agarwal}},\ }\emph {\bibinfo {title} {3{D} {D}irac
  Plasmons in the Type-{II} {D}irac Semimetal ${\mathrm{PtTe}}_{2}$}},\ \href
  {\doibase 10.1103/PhysRevLett.121.086804} {\bibfield  {journal} {\bibinfo
  {journal} {Phys. Rev. Lett.}\ }\textbf {\bibinfo {volume} {121}},\ \bibinfo
  {pages} {086804} (\bibinfo {year} {2018})}\BibitemShut {NoStop}%
\bibitem [{\citenamefont {Kim}\ \emph {et~al.}(2018)\citenamefont {Kim},
  \citenamefont {Ahn}, \citenamefont {Jung}, \citenamefont {Min}, \citenamefont
  {Ihm}, \citenamefont {Han},\ and\ \citenamefont
  {Kim}}]{kim.prm.2.104203.2018}%
  \BibitemOpen
  \bibfield  {author} {\bibinfo {author} {\bibfnamefont {D.}~\bibnamefont
  {Kim}}, \bibinfo {author} {\bibfnamefont {S.}~\bibnamefont {Ahn}}, \bibinfo
  {author} {\bibfnamefont {J.~H.}\ \bibnamefont {Jung}}, \bibinfo {author}
  {\bibfnamefont {H.}~\bibnamefont {Min}}, \bibinfo {author} {\bibfnamefont
  {J.}~\bibnamefont {Ihm}}, \bibinfo {author} {\bibfnamefont {J.~H.}\
  \bibnamefont {Han}}, \ and\ \bibinfo {author} {\bibfnamefont
  {Y.}~\bibnamefont {Kim}},\ }\emph {\bibinfo {title} {Type-{II} Dirac line
  node in strained ${\mathrm{Na}}_{3}\mathrm{N}$}},\ \href {\doibase
  10.1103/PhysRevMaterials.2.104203} {\bibfield  {journal} {\bibinfo  {journal}
  {Phys. Rev. Mater.}\ }\textbf {\bibinfo {volume} {2}},\ \bibinfo {pages}
  {104203} (\bibinfo {year} {2018})}\BibitemShut {NoStop}%
\bibitem [{\citenamefont {Wang}\ \emph {et~al.}(2018)\citenamefont {Wang},
  \citenamefont {Gao}, \citenamefont {Lu}, \citenamefont {Xie}, \citenamefont
  {Ge}, \citenamefont {Zhao}, \citenamefont {Zhang},\ and\ \citenamefont
  {Liu}}]{wang.prb.98.115164.2018}%
  \BibitemOpen
  \bibfield  {author} {\bibinfo {author} {\bibfnamefont {B.}~\bibnamefont
  {Wang}}, \bibinfo {author} {\bibfnamefont {H.}~\bibnamefont {Gao}}, \bibinfo
  {author} {\bibfnamefont {Q.}~\bibnamefont {Lu}}, \bibinfo {author}
  {\bibfnamefont {W.}~\bibnamefont {Xie}}, \bibinfo {author} {\bibfnamefont
  {Y.}~\bibnamefont {Ge}}, \bibinfo {author} {\bibfnamefont {Y.-H.}\
  \bibnamefont {Zhao}}, \bibinfo {author} {\bibfnamefont {K.}~\bibnamefont
  {Zhang}}, \ and\ \bibinfo {author} {\bibfnamefont {Y.}~\bibnamefont {Liu}},\
  }\emph {\bibinfo {title} {Type-{I} and type-{II} nodal lines coexistence in
  the antiferromagnetic monolayer ${\mathrm{CrAs}}_{2}$}},\ \href {\doibase
  10.1103/PhysRevB.98.115164} {\bibfield  {journal} {\bibinfo  {journal} {Phys.
  Rev. B}\ }\textbf {\bibinfo {volume} {98}},\ \bibinfo {pages} {115164}
  (\bibinfo {year} {2018})}\BibitemShut {NoStop}%
\bibitem [{\citenamefont {Lin}\ \emph {et~al.}(2017)\citenamefont {Lin},
  \citenamefont {Hu}, \citenamefont {Chen}, \citenamefont {Lee},\ and\
  \citenamefont {Zhang}}]{lin.prb.96.075438.2017}%
  \BibitemOpen
  \bibfield  {author} {\bibinfo {author} {\bibfnamefont {J.~Y.}\ \bibnamefont
  {Lin}}, \bibinfo {author} {\bibfnamefont {N.~C.}\ \bibnamefont {Hu}},
  \bibinfo {author} {\bibfnamefont {Y.~J.}\ \bibnamefont {Chen}}, \bibinfo
  {author} {\bibfnamefont {C.~H.}\ \bibnamefont {Lee}}, \ and\ \bibinfo
  {author} {\bibfnamefont {X.}~\bibnamefont {Zhang}},\ }\emph {\bibinfo {title}
  {Line nodes, {D}irac points, and {L}ifshitz transition in two-dimensional
  nonsymmorphic photonic crystals}},\ \href {\doibase
  10.1103/PhysRevB.96.075438} {\bibfield  {journal} {\bibinfo  {journal} {Phys.
  Rev. B}\ }\textbf {\bibinfo {volume} {96}},\ \bibinfo {pages} {075438}
  (\bibinfo {year} {2017})}\BibitemShut {NoStop}%
\bibitem [{\citenamefont {Pyrialakos}\ \emph {et~al.}(2017)\citenamefont
  {Pyrialakos}, \citenamefont {Nye}, \citenamefont {Kantartzis},\ and\
  \citenamefont {Christodoulides}}]{pyrialakos.prl.119.113901.2017}%
  \BibitemOpen
  \bibfield  {author} {\bibinfo {author} {\bibfnamefont {G.~G.}\ \bibnamefont
  {Pyrialakos}}, \bibinfo {author} {\bibfnamefont {N.~S.}\ \bibnamefont {Nye}},
  \bibinfo {author} {\bibfnamefont {N.~V.}\ \bibnamefont {Kantartzis}}, \ and\
  \bibinfo {author} {\bibfnamefont {D.~N.}\ \bibnamefont {Christodoulides}},\
  }\emph {\bibinfo {title} {Emergence of Type-{II} {D}irac Points in
  Graphynelike Photonic Lattices}},\ \href {\doibase
  10.1103/PhysRevLett.119.113901} {\bibfield  {journal} {\bibinfo  {journal}
  {Phys. Rev. Lett.}\ }\textbf {\bibinfo {volume} {119}},\ \bibinfo {pages}
  {113901} (\bibinfo {year} {2017})}\BibitemShut {NoStop}%
\bibitem [{\citenamefont {Wang}\ \emph {et~al.}(2017)\citenamefont {Wang},
  \citenamefont {Chen}, \citenamefont {Hang}, \citenamefont {Kee},\ and\
  \citenamefont {Jiang}}]{wang.npjqm.2.54.2017}%
  \BibitemOpen
  \bibfield  {author} {\bibinfo {author} {\bibfnamefont {H.-X.}\ \bibnamefont
  {Wang}}, \bibinfo {author} {\bibfnamefont {Y.}~\bibnamefont {Chen}}, \bibinfo
  {author} {\bibfnamefont {Z.~H.}\ \bibnamefont {Hang}}, \bibinfo {author}
  {\bibfnamefont {H.-Y.}\ \bibnamefont {Kee}}, \ and\ \bibinfo {author}
  {\bibfnamefont {J.-H.}\ \bibnamefont {Jiang}},\ }\emph {\bibinfo {title}
  {Type-{II} {D}irac photons}},\ \href
  {https://doi.org/10.1038/s41535-017-0058-z} {\bibfield  {journal} {\bibinfo
  {journal} {npj Quant. Mater.}\ }\textbf {\bibinfo {volume} {2}},\ \bibinfo
  {pages} {54} (\bibinfo {year} {2017})}\BibitemShut {NoStop}%
\bibitem [{\citenamefont {Hu}\ \emph {et~al.}(2018)\citenamefont {Hu},
  \citenamefont {Li}, \citenamefont {Tong}, \citenamefont {Wu}, \citenamefont
  {Xia}, \citenamefont {Wang}, \citenamefont {Li}, \citenamefont {Huang},
  \citenamefont {Wang}, \citenamefont {Hou}, \citenamefont {Chan},\ and\
  \citenamefont {Wen}}]{hu.prl.121.024301.2018}%
  \BibitemOpen
  \bibfield  {author} {\bibinfo {author} {\bibfnamefont {C.}~\bibnamefont
  {Hu}}, \bibinfo {author} {\bibfnamefont {Z.}~\bibnamefont {Li}}, \bibinfo
  {author} {\bibfnamefont {R.}~\bibnamefont {Tong}}, \bibinfo {author}
  {\bibfnamefont {X.}~\bibnamefont {Wu}}, \bibinfo {author} {\bibfnamefont
  {Z.}~\bibnamefont {Xia}}, \bibinfo {author} {\bibfnamefont {L.}~\bibnamefont
  {Wang}}, \bibinfo {author} {\bibfnamefont {S.}~\bibnamefont {Li}}, \bibinfo
  {author} {\bibfnamefont {Y.}~\bibnamefont {Huang}}, \bibinfo {author}
  {\bibfnamefont {S.}~\bibnamefont {Wang}}, \bibinfo {author} {\bibfnamefont
  {B.}~\bibnamefont {Hou}}, \bibinfo {author} {\bibfnamefont {C.~T.}\
  \bibnamefont {Chan}}, \ and\ \bibinfo {author} {\bibfnamefont
  {W.}~\bibnamefont {Wen}},\ }\emph {\bibinfo {title} {Type-{II} {D}irac
  Photons at Metasurfaces}},\ \href {\doibase 10.1103/PhysRevLett.121.024301}
  {\bibfield  {journal} {\bibinfo  {journal} {Phys. Rev. Lett.}\ }\textbf
  {\bibinfo {volume} {121}},\ \bibinfo {pages} {024301} (\bibinfo {year}
  {2018})}\BibitemShut {NoStop}%
\bibitem [{\citenamefont {Mann}\ \emph {et~al.}(2018)\citenamefont {Mann},
  \citenamefont {Sturges}, \citenamefont {Weick}, \citenamefont {Barnes},\ and\
  \citenamefont {Mariani}}]{mann.nc.9.2194.2018}%
  \BibitemOpen
  \bibfield  {author} {\bibinfo {author} {\bibfnamefont {C.-R.}\ \bibnamefont
  {Mann}}, \bibinfo {author} {\bibfnamefont {T.~J.}\ \bibnamefont {Sturges}},
  \bibinfo {author} {\bibfnamefont {G.}~\bibnamefont {Weick}}, \bibinfo
  {author} {\bibfnamefont {W.~L.}\ \bibnamefont {Barnes}}, \ and\ \bibinfo
  {author} {\bibfnamefont {E.}~\bibnamefont {Mariani}},\ }\emph {\bibinfo
  {title} {Manipulating type-{I} and type-{II} {D}irac polaritons in
  cavity-embedded honeycomb metasurfaces}},\ \href
  {https://doi.org/10.1038/s41467-018-03982-7} {\bibfield  {journal} {\bibinfo
  {journal} {Nat. Commun.}\ }\textbf {\bibinfo {volume} {9}},\ \bibinfo {pages}
  {2194} (\bibinfo {year} {2018})}\BibitemShut {NoStop}%
\bibitem [{\citenamefont {Li}\ \emph {et~al.}(2017)\citenamefont {Li},
  \citenamefont {Yu}, \citenamefont {Liu}, \citenamefont {Guan}, \citenamefont
  {Wang}, \citenamefont {Zhang}, \citenamefont {Yao},\ and\ \citenamefont
  {Yang}}]{li.prb.96.081106.2017}%
  \BibitemOpen
  \bibfield  {author} {\bibinfo {author} {\bibfnamefont {S.}~\bibnamefont
  {Li}}, \bibinfo {author} {\bibfnamefont {Z.-M.}\ \bibnamefont {Yu}}, \bibinfo
  {author} {\bibfnamefont {Y.}~\bibnamefont {Liu}}, \bibinfo {author}
  {\bibfnamefont {S.}~\bibnamefont {Guan}}, \bibinfo {author} {\bibfnamefont
  {S.-S.}\ \bibnamefont {Wang}}, \bibinfo {author} {\bibfnamefont
  {X.}~\bibnamefont {Zhang}}, \bibinfo {author} {\bibfnamefont
  {Y.}~\bibnamefont {Yao}}, \ and\ \bibinfo {author} {\bibfnamefont {S.~A.}\
  \bibnamefont {Yang}},\ }\emph {\bibinfo {title} {Type-{II} nodal loops:
  Theory and material realization}},\ \href {\doibase
  10.1103/PhysRevB.96.081106} {\bibfield  {journal} {\bibinfo  {journal} {Phys.
  Rev. B}\ }\textbf {\bibinfo {volume} {96}},\ \bibinfo {pages} {081106}
  (\bibinfo {year} {2017})}\BibitemShut {NoStop}%
\bibitem [{\citenamefont {Katsnelson}\ \emph {et~al.}(2006)\citenamefont
  {Katsnelson}, \citenamefont {Novoselov},\ and\ \citenamefont
  {Geim}}]{katsnelson.np.2.620.2006}%
  \BibitemOpen
  \bibfield  {author} {\bibinfo {author} {\bibfnamefont {M.~I.}\ \bibnamefont
  {Katsnelson}}, \bibinfo {author} {\bibfnamefont {K.~S.}\ \bibnamefont
  {Novoselov}}, \ and\ \bibinfo {author} {\bibfnamefont {A.~K.}\ \bibnamefont
  {Geim}},\ }\emph {\bibinfo {title} {Chiral tunnelling and the {K}lein paradox
  in graphene}},\ \href {https://doi.org/10.1038/nphys384} {\bibfield
  {journal} {\bibinfo  {journal} {Nat. Phys.}\ }\textbf {\bibinfo {volume}
  {2}},\ \bibinfo {pages} {620} (\bibinfo {year} {2006})}\BibitemShut {NoStop}%
\bibitem [{\citenamefont {Bahat-Treidel}\ \emph {et~al.}(2010)\citenamefont
  {Bahat-Treidel}, \citenamefont {Peleg}, \citenamefont {Grobman},
  \citenamefont {Shapira}, \citenamefont {Segev},\ and\ \citenamefont
  {Pereg-Barnea}}]{bahat-treidel.prl.104.063901.2010}%
  \BibitemOpen
  \bibfield  {author} {\bibinfo {author} {\bibfnamefont {O.}~\bibnamefont
  {Bahat-Treidel}}, \bibinfo {author} {\bibfnamefont {O.}~\bibnamefont
  {Peleg}}, \bibinfo {author} {\bibfnamefont {M.}~\bibnamefont {Grobman}},
  \bibinfo {author} {\bibfnamefont {N.}~\bibnamefont {Shapira}}, \bibinfo
  {author} {\bibfnamefont {M.}~\bibnamefont {Segev}}, \ and\ \bibinfo {author}
  {\bibfnamefont {T.}~\bibnamefont {Pereg-Barnea}},\ }\emph {\bibinfo {title}
  {Klein Tunneling in Deformed Honeycomb Lattices}},\ \href {\doibase
  10.1103/PhysRevLett.104.063901} {\bibfield  {journal} {\bibinfo  {journal}
  {Phys. Rev. Lett.}\ }\textbf {\bibinfo {volume} {104}},\ \bibinfo {pages}
  {063901} (\bibinfo {year} {2010})}\BibitemShut {NoStop}%
\bibitem [{\citenamefont {Bahat-Treidel}\ and\ \citenamefont
  {Segev}(2011)}]{bahat-treidel.pra.84.021802.2011}%
  \BibitemOpen
  \bibfield  {author} {\bibinfo {author} {\bibfnamefont {O.}~\bibnamefont
  {Bahat-Treidel}}\ and\ \bibinfo {author} {\bibfnamefont {M.}~\bibnamefont
  {Segev}},\ }\emph {\bibinfo {title} {Nonlinear wave dynamics in honeycomb
  lattices}},\ \href {\doibase 10.1103/PhysRevA.84.021802} {\bibfield
  {journal} {\bibinfo  {journal} {Phys. Rev. A}\ }\textbf {\bibinfo {volume}
  {84}},\ \bibinfo {pages} {021802} (\bibinfo {year} {2011})}\BibitemShut
  {NoStop}%
\bibitem [{\citenamefont {Longhi}(2010)}]{longhi.prb.81.075102.2010}%
  \BibitemOpen
  \bibfield  {author} {\bibinfo {author} {\bibfnamefont {S.}~\bibnamefont
  {Longhi}},\ }\emph {\bibinfo {title} {Klein tunneling in binary photonic
  superlattices}},\ \href {\doibase 10.1103/PhysRevB.81.075102} {\bibfield
  {journal} {\bibinfo  {journal} {Phys. Rev. B}\ }\textbf {\bibinfo {volume}
  {81}},\ \bibinfo {pages} {075102} (\bibinfo {year} {2010})}\BibitemShut
  {NoStop}%
\bibitem [{\citenamefont {Dreisow}\ \emph {et~al.}(2012)\citenamefont
  {Dreisow}, \citenamefont {Keil}, \citenamefont {T\"unnermann}, \citenamefont
  {Nolte}, \citenamefont {Longhi},\ and\ \citenamefont
  {Szameit}}]{dreisow.epl.97.10008.2012}%
  \BibitemOpen
  \bibfield  {author} {\bibinfo {author} {\bibfnamefont {F.}~\bibnamefont
  {Dreisow}}, \bibinfo {author} {\bibfnamefont {R.}~\bibnamefont {Keil}},
  \bibinfo {author} {\bibfnamefont {A.}~\bibnamefont {T\"unnermann}}, \bibinfo
  {author} {\bibfnamefont {S.}~\bibnamefont {Nolte}}, \bibinfo {author}
  {\bibfnamefont {S.}~\bibnamefont {Longhi}}, \ and\ \bibinfo {author}
  {\bibfnamefont {A.}~\bibnamefont {Szameit}},\ }\emph {\bibinfo {title} {Klein
  tunneling of light in waveguide superlattices}},\ \href {\doibase
  10.1209/0295-5075/97/10008} {\bibfield  {journal} {\bibinfo  {journal} {EPL}\
  }\textbf {\bibinfo {volume} {97}},\ \bibinfo {pages} {10008} (\bibinfo {year}
  {2012})}\BibitemShut {NoStop}%
\bibitem [{\citenamefont {Sun}\ \emph {et~al.}(2017)\citenamefont {Sun},
  \citenamefont {Gao},\ and\ \citenamefont {Yang}}]{sun.sr.7.9678.2017}%
  \BibitemOpen
  \bibfield  {author} {\bibinfo {author} {\bibfnamefont {L.}~\bibnamefont
  {Sun}}, \bibinfo {author} {\bibfnamefont {J.}~\bibnamefont {Gao}}, \ and\
  \bibinfo {author} {\bibfnamefont {X.}~\bibnamefont {Yang}},\ }\emph {\bibinfo
  {title} {Klein tunneling near the {D}irac points in metal-dielectric
  multilayer metamaterials}},\ \href
  {https://doi.org/10.1038/s41598-017-09899-3} {\bibfield  {journal} {\bibinfo
  {journal} {Sci. Rep.}\ }\textbf {\bibinfo {volume} {7}},\ \bibinfo {pages}
  {9678} (\bibinfo {year} {2017})}\BibitemShut {NoStop}%
\bibitem [{\citenamefont {Fang}\ \emph {et~al.}(2019)\citenamefont {Fang},
  \citenamefont {Zhang}, \citenamefont {Louie},\ and\ \citenamefont
  {Chan}}]{fang.research.2019.3054062.2019}%
  \BibitemOpen
  \bibfield  {author} {\bibinfo {author} {\bibfnamefont {A.}~\bibnamefont
  {Fang}}, \bibinfo {author} {\bibfnamefont {Z.~Q.}\ \bibnamefont {Zhang}},
  \bibinfo {author} {\bibfnamefont {S.~G.}\ \bibnamefont {Louie}}, \ and\
  \bibinfo {author} {\bibfnamefont {C.~T.}\ \bibnamefont {Chan}},\ }\emph
  {\bibinfo {title} {Pseudospin-1 Physics of Photonic Crystals}},\ \href
  {\doibase 10.1155/2019/3054062} {\bibfield  {journal} {\bibinfo  {journal}
  {Research}\ }\textbf {\bibinfo {volume} {2019}},\ \bibinfo {pages} {3054062}
  (\bibinfo {year} {2019})}\BibitemShut {NoStop}%
\bibitem [{\citenamefont {Li}\ \emph {et~al.}(2018)\citenamefont {Li},
  \citenamefont {Ye}, \citenamefont {Chen}, \citenamefont {Kartashov},
  \citenamefont {Ferrando}, \citenamefont {Torner},\ and\ \citenamefont
  {Skryabin}}]{li.prb.97.081103.2018}%
  \BibitemOpen
  \bibfield  {author} {\bibinfo {author} {\bibfnamefont {C.}~\bibnamefont
  {Li}}, \bibinfo {author} {\bibfnamefont {F.}~\bibnamefont {Ye}}, \bibinfo
  {author} {\bibfnamefont {X.}~\bibnamefont {Chen}}, \bibinfo {author}
  {\bibfnamefont {Y.~V.}\ \bibnamefont {Kartashov}}, \bibinfo {author}
  {\bibfnamefont {A.}~\bibnamefont {Ferrando}}, \bibinfo {author}
  {\bibfnamefont {L.}~\bibnamefont {Torner}}, \ and\ \bibinfo {author}
  {\bibfnamefont {D.~V.}\ \bibnamefont {Skryabin}},\ }\emph {\bibinfo {title}
  {Lieb polariton topological insulators}},\ \href {\doibase
  10.1103/PhysRevB.97.081103} {\bibfield  {journal} {\bibinfo  {journal} {Phys.
  Rev. B}\ }\textbf {\bibinfo {volume} {97}},\ \bibinfo {pages} {081103}
  (\bibinfo {year} {2018})}\BibitemShut {NoStop}%
\bibitem [{\citenamefont {Rechtsman}\ \emph
  {et~al.}(2013{\natexlab{c}})\citenamefont {Rechtsman}, \citenamefont
  {Zeuner}, \citenamefont {Plotnik}, \citenamefont {Lumer}, \citenamefont
  {Podolsky}, \citenamefont {Dreisow}, \citenamefont {Nolte}, \citenamefont
  {Segev},\ and\ \citenamefont {Szameit}}]{rechtsman.nature.496.196.2013}%
  \BibitemOpen
  \bibfield  {author} {\bibinfo {author} {\bibfnamefont {M.~C.}\ \bibnamefont
  {Rechtsman}}, \bibinfo {author} {\bibfnamefont {J.~M.}\ \bibnamefont
  {Zeuner}}, \bibinfo {author} {\bibfnamefont {Y.}~\bibnamefont {Plotnik}},
  \bibinfo {author} {\bibfnamefont {Y.}~\bibnamefont {Lumer}}, \bibinfo
  {author} {\bibfnamefont {D.}~\bibnamefont {Podolsky}}, \bibinfo {author}
  {\bibfnamefont {F.}~\bibnamefont {Dreisow}}, \bibinfo {author} {\bibfnamefont
  {S.}~\bibnamefont {Nolte}}, \bibinfo {author} {\bibfnamefont
  {M.}~\bibnamefont {Segev}}, \ and\ \bibinfo {author} {\bibfnamefont
  {A.}~\bibnamefont {Szameit}},\ }\emph {\bibinfo {title} {Photonic Floquet
  topological insulators}},\ \href {https://doi.org/10.1038/nature12066}
  {\bibfield  {journal} {\bibinfo  {journal} {Nature}\ }\textbf {\bibinfo
  {volume} {496}},\ \bibinfo {pages} {196} (\bibinfo {year}
  {2013}{\natexlab{c}})}\BibitemShut {NoStop}%
\bibitem [{\citenamefont {St\"utzer}\ \emph {et~al.}(2018)\citenamefont
  {St\"utzer}, \citenamefont {Plotnik}, \citenamefont {Lumer}, \citenamefont
  {Titum}, \citenamefont {Lindner}, \citenamefont {Segev}, \citenamefont
  {Rechtsman},\ and\ \citenamefont {Szameit}}]{stuzer.nature.560.461.2019}%
  \BibitemOpen
  \bibfield  {author} {\bibinfo {author} {\bibfnamefont {S.}~\bibnamefont
  {St\"utzer}}, \bibinfo {author} {\bibfnamefont {Y.}~\bibnamefont {Plotnik}},
  \bibinfo {author} {\bibfnamefont {Y.}~\bibnamefont {Lumer}}, \bibinfo
  {author} {\bibfnamefont {P.}~\bibnamefont {Titum}}, \bibinfo {author}
  {\bibfnamefont {N.~H.}\ \bibnamefont {Lindner}}, \bibinfo {author}
  {\bibfnamefont {M.}~\bibnamefont {Segev}}, \bibinfo {author} {\bibfnamefont
  {M.~C.}\ \bibnamefont {Rechtsman}}, \ and\ \bibinfo {author} {\bibfnamefont
  {A.}~\bibnamefont {Szameit}},\ }\emph {\bibinfo {title} {Photonic topological
  Anderson insulators}},\ \href {https://doi.org/10.1038/s41586-018-0418-2}
  {\bibfield  {journal} {\bibinfo  {journal} {Nature}\ }\textbf {\bibinfo
  {volume} {560}},\ \bibinfo {pages} {461} (\bibinfo {year}
  {2018})}\BibitemShut {NoStop}%
\bibitem [{\citenamefont {Lustig}\ \emph {et~al.}(2019)\citenamefont {Lustig},
  \citenamefont {Weimann}, \citenamefont {Plotnik}, \citenamefont {Lumer},
  \citenamefont {Bandres}, \citenamefont {Szameit},\ and\ \citenamefont
  {Segev}}]{lustig.nature.567.356.2019}%
  \BibitemOpen
  \bibfield  {author} {\bibinfo {author} {\bibfnamefont {E.}~\bibnamefont
  {Lustig}}, \bibinfo {author} {\bibfnamefont {S.}~\bibnamefont {Weimann}},
  \bibinfo {author} {\bibfnamefont {Y.}~\bibnamefont {Plotnik}}, \bibinfo
  {author} {\bibfnamefont {Y.}~\bibnamefont {Lumer}}, \bibinfo {author}
  {\bibfnamefont {M.~A.}\ \bibnamefont {Bandres}}, \bibinfo {author}
  {\bibfnamefont {A.}~\bibnamefont {Szameit}}, \ and\ \bibinfo {author}
  {\bibfnamefont {M.}~\bibnamefont {Segev}},\ }\emph {\bibinfo {title}
  {Photonic topological insulator in synthetic dimensions}},\ \href
  {https://doi.org/10.1038/s41586-019-0943-7} {\bibfield  {journal} {\bibinfo
  {journal} {Nature}\ }\textbf {\bibinfo {volume} {567}},\ \bibinfo {pages}
  {356} (\bibinfo {year} {2019})}\BibitemShut {NoStop}%
\bibitem [{\citenamefont {Wang}\ \emph {et~al.}(2020)\citenamefont {Wang},
  \citenamefont {Zheng}, \citenamefont {Chen}, \citenamefont {Huang},
  \citenamefont {Kartashov}, \citenamefont {Torner}, \citenamefont {Konotop},\
  and\ \citenamefont {Ye}}]{wang.nature.577.42.2020}%
  \BibitemOpen
  \bibfield  {author} {\bibinfo {author} {\bibfnamefont {P.}~\bibnamefont
  {Wang}}, \bibinfo {author} {\bibfnamefont {Y.}~\bibnamefont {Zheng}},
  \bibinfo {author} {\bibfnamefont {X.}~\bibnamefont {Chen}}, \bibinfo {author}
  {\bibfnamefont {C.}~\bibnamefont {Huang}}, \bibinfo {author} {\bibfnamefont
  {Y.~V.}\ \bibnamefont {Kartashov}}, \bibinfo {author} {\bibfnamefont
  {L.}~\bibnamefont {Torner}}, \bibinfo {author} {\bibfnamefont {V.~V.}\
  \bibnamefont {Konotop}}, \ and\ \bibinfo {author} {\bibfnamefont
  {F.}~\bibnamefont {Ye}},\ }\emph {\bibinfo {title} {Localization and
  delocalization of light in photonic moir\'e lattices}},\ \href
  {https://doi.org/10.1038/s41586-019-1851-6} {\bibfield  {journal} {\bibinfo
  {journal} {Nature}\ }\textbf {\bibinfo {volume} {577}},\ \bibinfo {pages}
  {42} (\bibinfo {year} {2020})}\BibitemShut {NoStop}%
\bibitem [{\citenamefont {Tarruell}\ \emph {et~al.}(2012)\citenamefont
  {Tarruell}, \citenamefont {Greif}, \citenamefont {Uehlinger}, \citenamefont
  {Jotzu},\ and\ \citenamefont {Esslinger}}]{tarruell.nature.483.302.2012}%
  \BibitemOpen
  \bibfield  {author} {\bibinfo {author} {\bibfnamefont {L.}~\bibnamefont
  {Tarruell}}, \bibinfo {author} {\bibfnamefont {D.}~\bibnamefont {Greif}},
  \bibinfo {author} {\bibfnamefont {T.}~\bibnamefont {Uehlinger}}, \bibinfo
  {author} {\bibfnamefont {G.}~\bibnamefont {Jotzu}}, \ and\ \bibinfo {author}
  {\bibfnamefont {T.}~\bibnamefont {Esslinger}},\ }\emph {\bibinfo {title}
  {Creating, moving and merging {D}irac points with a {F}ermi gas in a tunable
  honeycomb lattice}},\ \href {https://doi.org/10.1038/nature10871} {\bibfield
  {journal} {\bibinfo  {journal} {Nature}\ }\textbf {\bibinfo {volume} {483}},\
  \bibinfo {pages} {302} (\bibinfo {year} {2012})}\BibitemShut {NoStop}%
\bibitem [{\citenamefont {Bartal}\ \emph {et~al.}(2005)\citenamefont {Bartal},
  \citenamefont {Cohen}, \citenamefont {Buljan}, \citenamefont {Fleischer},
  \citenamefont {Manela},\ and\ \citenamefont
  {Segev}}]{bartal.prl.94.163902.2005}%
  \BibitemOpen
  \bibfield  {author} {\bibinfo {author} {\bibfnamefont {G.}~\bibnamefont
  {Bartal}}, \bibinfo {author} {\bibfnamefont {O.}~\bibnamefont {Cohen}},
  \bibinfo {author} {\bibfnamefont {H.}~\bibnamefont {Buljan}}, \bibinfo
  {author} {\bibfnamefont {J.~W.}\ \bibnamefont {Fleischer}}, \bibinfo {author}
  {\bibfnamefont {O.}~\bibnamefont {Manela}}, \ and\ \bibinfo {author}
  {\bibfnamefont {M.}~\bibnamefont {Segev}},\ }\emph {\bibinfo {title}
  {Brillouin Zone Spectroscopy of Nonlinear Photonic Lattices}},\ \href
  {\doibase 10.1103/PhysRevLett.94.163902} {\bibfield  {journal} {\bibinfo
  {journal} {Phys. Rev. Lett.}\ }\textbf {\bibinfo {volume} {94}},\ \bibinfo
  {pages} {163902} (\bibinfo {year} {2005})}\BibitemShut {NoStop}%
\bibitem [{\citenamefont {Peleg}\ \emph {et~al.}(2007)\citenamefont {Peleg},
  \citenamefont {Bartal}, \citenamefont {Freedman}, \citenamefont {Manela},
  \citenamefont {Segev},\ and\ \citenamefont
  {Christodoulides}}]{peleg.prl.98.103901.2007}%
  \BibitemOpen
  \bibfield  {author} {\bibinfo {author} {\bibfnamefont {O.}~\bibnamefont
  {Peleg}}, \bibinfo {author} {\bibfnamefont {G.}~\bibnamefont {Bartal}},
  \bibinfo {author} {\bibfnamefont {B.}~\bibnamefont {Freedman}}, \bibinfo
  {author} {\bibfnamefont {O.}~\bibnamefont {Manela}}, \bibinfo {author}
  {\bibfnamefont {M.}~\bibnamefont {Segev}}, \ and\ \bibinfo {author}
  {\bibfnamefont {D.~N.}\ \bibnamefont {Christodoulides}},\ }\emph {\bibinfo
  {title} {Conical Diffraction and Gap Solitons in Honeycomb Photonic
  Lattices}},\ \href {\doibase 10.1103/PhysRevLett.98.103901} {\bibfield
  {journal} {\bibinfo  {journal} {Phys. Rev. Lett.}\ }\textbf {\bibinfo
  {volume} {98}},\ \bibinfo {pages} {103901} (\bibinfo {year}
  {2007})}\BibitemShut {NoStop}%
\bibitem [{\citenamefont {Ablowitz}\ \emph {et~al.}(2009)\citenamefont
  {Ablowitz}, \citenamefont {Nixon},\ and\ \citenamefont
  {Zhu}}]{ablowitz.pra.79.053830.2009}%
  \BibitemOpen
  \bibfield  {author} {\bibinfo {author} {\bibfnamefont {M.~J.}\ \bibnamefont
  {Ablowitz}}, \bibinfo {author} {\bibfnamefont {S.~D.}\ \bibnamefont {Nixon}},
  \ and\ \bibinfo {author} {\bibfnamefont {Y.}~\bibnamefont {Zhu}},\ }\emph
  {\bibinfo {title} {Conical diffraction in honeycomb lattices}},\ \href
  {\doibase 10.1103/PhysRevA.79.053830} {\bibfield  {journal} {\bibinfo
  {journal} {Phys. Rev. A}\ }\textbf {\bibinfo {volume} {79}},\ \bibinfo
  {pages} {053830} (\bibinfo {year} {2009})}\BibitemShut {NoStop}%
\bibitem [{\citenamefont {Wu}\ \emph {et~al.}(2020)\citenamefont {Wu},
  \citenamefont {Li}, \citenamefont {Zhang}, \citenamefont {Xiang},
  \citenamefont {Tian}, \citenamefont {Huang}, \citenamefont {Wang},
  \citenamefont {Hou}, \citenamefont {Chan},\ and\ \citenamefont
  {Wen}}]{wu.prl.124.075501.2020}%
  \BibitemOpen
  \bibfield  {author} {\bibinfo {author} {\bibfnamefont {X.}~\bibnamefont
  {Wu}}, \bibinfo {author} {\bibfnamefont {X.}~\bibnamefont {Li}}, \bibinfo
  {author} {\bibfnamefont {R.-Y.}\ \bibnamefont {Zhang}}, \bibinfo {author}
  {\bibfnamefont {X.}~\bibnamefont {Xiang}}, \bibinfo {author} {\bibfnamefont
  {J.}~\bibnamefont {Tian}}, \bibinfo {author} {\bibfnamefont {Y.}~\bibnamefont
  {Huang}}, \bibinfo {author} {\bibfnamefont {S.}~\bibnamefont {Wang}},
  \bibinfo {author} {\bibfnamefont {B.}~\bibnamefont {Hou}}, \bibinfo {author}
  {\bibfnamefont {C.~T.}\ \bibnamefont {Chan}}, \ and\ \bibinfo {author}
  {\bibfnamefont {W.}~\bibnamefont {Wen}},\ }\emph {\bibinfo {title}
  {Deterministic Scheme for Two-Dimensional Type-{II D}irac Points and
  Experimental Realization in Acoustics}},\ \href {\doibase
  10.1103/PhysRevLett.124.075501} {\bibfield  {journal} {\bibinfo  {journal}
  {Phys. Rev. Lett.}\ }\textbf {\bibinfo {volume} {124}},\ \bibinfo {pages}
  {075501} (\bibinfo {year} {2020})}\BibitemShut {NoStop}%
\bibitem [{\citenamefont {Pyrialakos}\ \emph {et~al.}(2020)\citenamefont
  {Pyrialakos}, \citenamefont {Schmitt}, \citenamefont {Nye}, \citenamefont
  {Heinrich}, \citenamefont {Kantartzis}, \citenamefont {Szameit},\ and\
  \citenamefont {Christodoulides}}]{pyrialakos.nc.11.2074.2020}%
  \BibitemOpen
  \bibfield  {author} {\bibinfo {author} {\bibfnamefont {G.~G.}\ \bibnamefont
  {Pyrialakos}}, \bibinfo {author} {\bibfnamefont {N.}~\bibnamefont {Schmitt}},
  \bibinfo {author} {\bibfnamefont {N.~S.}\ \bibnamefont {Nye}}, \bibinfo
  {author} {\bibfnamefont {M.}~\bibnamefont {Heinrich}}, \bibinfo {author}
  {\bibfnamefont {N.~V.}\ \bibnamefont {Kantartzis}}, \bibinfo {author}
  {\bibfnamefont {A.}~\bibnamefont {Szameit}}, \ and\ \bibinfo {author}
  {\bibfnamefont {D.~N.}\ \bibnamefont {Christodoulides}},\ }\emph {\bibinfo
  {title} {Symmetry-controlled edge states in the type-{II} phase of {D}irac
  photonic lattices}},\ \href {https://doi.org/10.1038/s41467-020-15952-z}
  {\bibfield  {journal} {\bibinfo  {journal} {Nat. Commun.}\ }\textbf {\bibinfo
  {volume} {11}},\ \bibinfo {pages} {2074} (\bibinfo {year}
  {2020})}\BibitemShut {NoStop}%
\end{thebibliography}%

\end{document}